\documentclass[11pt,a4paper]{article}
\pdfoutput=1
\usepackage{jheppub}

\usepackage{multirow}
\usepackage{amsmath,amssymb}
\usepackage{graphicx}
\usepackage{color}
\usepackage{hyperref}
\usepackage{url}
\usepackage{slashed}
\usepackage{subfigure}
\usepackage[usenames,dvipsnames]{xcolor}

\title{Perturbative asymptotic safety and the graviton}
\author[a]{Nicolai Christiansen,}
\author[a]{Astrid Eichhorn,}
\author[a]{Aaron Held}
\emailAdd{n.christiansen@thphys.uni-heidelberg.de, a.eichhorn@thphys.uni-heidelberg.de, a.held@thphys.uni-heidelberg.de}

\affiliation[a]{Institut f\"ur Theoretische Physik, Universit\"at Heidelberg, Philosophenweg 16, 69120 Heidelberg, Germany}
\usepackage{amssymb}
\usepackage{slashed}
\usepackage{graphicx}
\usepackage{graphics}
\usepackage{epsfig}
\usepackage{color}
\usepackage{soul}
\usepackage{tabularx}
\usepackage{amsmath}
\usepackage{amsfonts}
\usepackage{amssymb}
\usepackage{hyperref}
\usepackage{microtype}

\newcommand{\be}{\begin{equation}}
\newcommand{\ee}{\end{equation}}
\newcommand{\bea}{\begin{eqnarray}}
\newcommand{\eea}{\end{eqnarray}}

\title{Is scale-invariance in gauge-Yukawa systems compatible with the graviton?
}

\abstract{
We explore whether perturbative interacting fixed points in matter 
systems can persist under the impact of quantum gravity.
We first focus on semi-simple gauge theories and show that the leading order gravity contribution evaluated within the functional Renormalization Group framework preserves the perturbative fixed-point structure in these models discovered in \cite{Esbensen:2015cjw}. 
We highlight that the quantum-gravity contribution alters the scaling dimension of the gauge coupling, such that the system exhibits an effective dimensional reduction. We secondly explore the effect of metric fluctuations on asymptotically safe gauge-Yukawa systems which feature an asymptotically safe fixed point \cite{Litim:2014uca}. The same effective dimensional reduction that takes effect in pure gauge theories also impacts gauge-Yukawa systems. There, it appears to lead to a split of the degenerate free fixed point into an interacting infrared attractive fixed point and a partially ultraviolet attractive free fixed point. The quantum-gravity induced infrared fixed point moves towards the asymptotically safe fixed point of the matter system, and annihilates it at a critical value of the gravity coupling. 
 Even after that fixed-point annihilation, graviton effects leave behind new partially interacting fixed points for the matter sector.
}

\begin{document}
\maketitle

\section{Introduction}
Asymptotic safety \cite{Weinberg:1980gg} generalizes the paradigm of asymptotic 
freedom \cite{Gross:1973id,Politzer:1973fx} and thereby provides ultraviolet 
(UV) completions for quantum field theories. 
It hinges on the existence of an interacting Renormalization Group (RG) fixed point. If the running couplings of a model approach such a fixed point with finitely many UV-attractive directions at high momentum scales, the model enters a scale-invariant regime, which allows one to ``zoom in" on arbitrarily small length scales while keeping all observables finite at all scales and dependent on just a finite number of free parameters. In high-energy physics, the special case of asymptotic freedom, where the RG fixed point is a free one, 
plays a distinctive role, as it underlies the UV structure of non-Abelian gauge theories.
 Asymptotic safety, where the fixed point features residual interactions, is 
not simple to achieve in Standard-Model like theories in four dimensions, as 
all couplings in the Standard Model, with the exception of the Higgs mass 
parameter, are dimensionless. 
To achieve a scale-invariant regime at finite interaction strength, one must therefore balance quantum fluctuations of different fields against each other. 
 On the other hand, in statistical physics, interacting RG fixed points are of course much more prevalent than free ones, 
showing that asymptotic safety is not just an exotic property 
of quantum field theories. 
Models away from their critical dimensionality, which can play a major role in statistical physics, feature dimensionful couplings. Then, interacting fixed points can arise from a balance between quantum and canonical scaling. Thus, manifold examples for interacting fixed points exist for models away from 
their critical dimensionality, e.g., in non-Abelian gauge theories in 
$d=4+\epsilon$ dimensions \cite{Peskin1980,Gies:2003ic}, in O(N) models with a 
``bound-state" field in $d=6-\epsilon$ dimensions 
\cite{Fei:2014yja,Gracey:2015tta,Eichhorn:2016hdi}, in the Gross-Neveu model in 
$d=2+\epsilon$ \cite{Gawedzki:1985ed,Hands:1992be,Braun:2010tt}.

A breakthrough was reached in \cite{Litim:2014uca}, where asymptotic safety in 
a particular class of gauge-Yukawa systems in four dimensions 
in the perturbative regime has been discovered. There, the interplay of quantum fluctuations of 
scalars, fermions and gauge fields allows to achieve a 
balance that leads to quantum scale-invariance in the UV \cite{Bond:2016dvk,Bond:2017sem},  interestingly 
without requiring supersymmetry \cite{Intriligator:2015xxa}. The corresponding 
models feature similar building blocks to the Standard Model.
Thus, the asymptotic-safety paradigm could potentially play a role in beyond-Standard-Model physics at scales below the Planck scale \cite{Pelaggi:2017wzr,Bond:2017wut},  with possible consequences for phenomenological questions \cite{Sannino:2014lxa,Nielsen:2015una,Rischke:2015mea,Bajc:2016efj}. On the other hand, there are compelling hints for asymptotic safety of gravity without \cite{ASgravity} and with \cite{ASgravitymatter}  matter beyond the Planck scale, for reviews, see \cite{ASreviews}. In this setting, it is an intriguing question whether asymptotic safety in pure matter systems can persist at intermediate scales, i.e., within the regime where gravity can be treated as an effective field theory.

In the following, we will explore leading-order quantum- gravity corrections on 
gauge-Yukawa systems. As a ``warm-up", we will first discuss  
quantum- gravity effects on the fixed-point structure of semi-simple gauge theories with fermions in Sec.~\ref{sec:semisim}, and then include scalars to explore the impact of quantum gravity on asymptotic safety in gauge-Yukawa systems in Sec.~\ref{sec:QGgaugeYukawa}.
 In particular, we explore whether the fixed-point structure in gauge-Yukawa systems, which features a perturbative, asymptotically safe fixed point in the Veneziano limit, survives the effect of gravity fluctuations.
 Breaking 
  with tradition, the answer to  the question paraphrased in the title, will 
tentatively be positive.

\section{Quantum-gravity corrections to the fixed-point structure in semi-simple gauge theories}\label{sec:semisim}

In this work, we consider a regime where the running of the 
gravitational coupling is governed by classical behavior, i.e., the 
dimensionfull Newton coupling is constant. Its dimensionless counterpart must 
therefore run as a function of the momentum scale $k$ and is given  in terms 
of 
the Planck scale $M_{\mathrm{Planck}}$ according to
\be
g = 
\frac{k^2}{M_{\rm Planck}^2}\, .\label{eq:runningg}
\ee
Although the 
gravitational coupling $g$ is small below the Planck scale, its contributions to the 
$\beta$-functions in the gauge and matter sector can be important. 
The leading-order quantum-gravity 
effects are linear in the Newton coupling. In our setting, well below the Planck 
scale,  we assume that it is sufficient to neglect all higher-order 
effects.  
In particular, this implies that the 
only relevant quantum-gravity effects are encoded in direct quantum-gravity 
contributions to the anomalous scaling of the matter couplings at their fixed 
points.
Specifically, the preservation of global symmetries by quantum gravity within the effective field theory regime, that we observe in all approximations that have been explored within the framework of the  functional RG, implies that the quantum-gravity contribution to the running of a gauge coupling $\alpha$ is of the form
\be
\beta_{\alpha}\Big|_{\rm QG} = -\mathcal{D}_{\alpha}^{\rm grav}\alpha\, g.\label{eq:betaalpha_schematic}
\ee
This contribution is universal in that it does not depend on the gauge group, i.e., quantum gravity is ``blind" to non-spacetime indices. Thus, the gravity-correction to Abelian and non-Abelian gauge theories is the same.  The critical piece of information is the sign of $\mathcal{D}_{\alpha}^{\rm grav}$. Within different schemes, this quantity has been calculated in the Abelian and non-Abelian case in \cite{EFTabelian,Daum:2009dn,Folkerts:2011jz,Harst:2011zx,Christiansen:2017gtg}.

 To evaluate the quantum-gravity contribution, we employ functional 
RG techniques \cite{Wetterich:1992yh,Morris:1993qb}. The 
functional RG allows us to probe the scale-dependence of a 
quantum field theory, introduced by a mass-like regularization scheme. 
Specifically, an infrared  (IR) regulator function, that acts as a mass-like cutoff 
term, is added to the action in the generating functional. Beta functions are 
then obtained from the Wetterich equation \cite{Wetterich:1992yh}, that encodes 
the change in the effective dynamics under a change of the cutoff scale. The 
universal one-loop terms of canonically dimensionless couplings are 
straightforward to reproduce with the Wetterich equation. Two-loop 
contributions can also be obtained, see, e.g., \cite{Papenbrock,Litim:2001ky}, 
although  the two-loop coefficient is not straightforward to obtain in the case of non-Abelian gauge theories 
\cite{Reuter:1993kw,Gies:2002af}. The Wetterich equation is an excellent method 
to explore interacting fixed points in systems with dimensionful couplings, 
such as the Wilson-Fisher fixed point in three dimensions. There, 
quantitatively precise values for the critical exponents can be obtained within 
this scheme \cite{Canet:2003qd,Litim:2010tt}.  Thus, the Wetterich equation is a well-suited tool to 
evaluate quantum-gravity contributions to running couplings. For reviews of the 
technique, see   
\cite{Berges:2000ew,Polonyi:2001se,Pawlowski:2005xe,Gies:2006wv,
Delamotte:2007pf,Rosten:2010vm,Braun:2011pp}.\\ 
Let us note that due to the dimensionful nature of the Newton coupling, quantum-gravity effects should not be expected to have universal beta functions, even at the one-loop level. This is of course not a problem, as running couplings do not directly correspond to observables. On the other hand, we check explicitly that our choice of gauge does not impact our results for the signs of the gravity-contributions in App.~\ref{sec:gaugedependence}. For the metric fluctuations, we work in a covariant gauge with gauge parameters $\alpha=0$, $\beta=1$ in the main part.
Our results are obtained with a Litim-type cutoff \cite{Litim:2001up} of the form $R_k = Z_k (k^2-p^2)\theta(k^2-p^2)$. 

The sign of the gravity contribution in Eq.~\eqref{eq:betaalpha_schematic} is 
critical to determine the impact of gravity on the gauge system. A negative 
sign, i.e.\,$\mathcal{D}_{\alpha}^{\rm grav}>0$, preserves the full 
fixed-point structure of semi-simple gauge theories as analyzed in 
\cite{Esbensen:2015cjw}. It is vital to note that the sign of the gravity 
contribution is gauge-independent, cf.~App.~\ref{sec:gaugedependence}, see also 
\cite{Folkerts:2011jz}.

 Here, we first explore the effect of the quantum-gravity correction on the fixed-point structure in semi-simple gauge theories.  The model features an SU($N$) gauge group with gauge coupling $\alpha_N$ and $M$ fundamental Weyl fermions as well as $M$ antifundamental fermions. An ${\rm SU}(2)_L$ subgroup of the global ${\rm SU}(M)_L\times {\rm SU}(M)_R$ global symmetry is gauged, and features the gauge coupling $\alpha_2$. At particular values of $(N,M)$, the system exhibits a fully IR-attractive interacting fixed point as well as two semi-interacting fixed points with one UV-attractive direction \cite{Esbensen:2015cjw}.
For the gravity-free beta functions, the corresponding results can be found in  App.~C1 of \cite{Esbensen:2015cjw}. Here, we will repeat them for clarity; our new result consists in the addition of the gravity contribution. 
The full beta functions read
\bea
\beta_{\alpha_N}&=& \beta_{\alpha_N}\Big|_{\mathrm{grav}}- \left( \frac{11}{2}N - \frac{2}{3}M\right)\frac{\alpha_N^2}{2\pi} - \left(\frac{17}{3}N^2- \left( \frac{5}{3}N + \frac{N^2-1}{2N}\right)M \right) \frac{\alpha_N^3}{(2\pi)^2} \nonumber\\
&{}&+\frac{3}{4}\alpha_2 \frac{\alpha_N^2}{(2\pi)^2}, \label{eq:betaalphaN}\\
\beta_{\alpha_2}&=& \beta_{\alpha_2}\Big|_{\mathrm{grav}}- \left( \frac{22}{3}- \frac{2}{3}N\right) \frac{\alpha_2^2}{2\pi} -\left(\frac{68}{3}- \frac{49}{12}N \right)\frac{\alpha_2^3}{(2\pi)^2} - \frac{1-N^2}{4} \alpha_N \frac{\alpha_2^2}{(2\pi)^2} ,\label{eq:betaalpha2}
\eea
where the gravity contributions to the running of the gauge 
couplings is independent of the gauge group and takes the form
\begin{equation}
\left. \beta_{\alpha_{N/2}}\right|_{\mathrm{grav}} = - g 
\frac{\alpha_{N/2}}{2\pi} \,. \label{eq:betagravalpha}
\end{equation}
Note that the gravity correction is linear in the gauge coupling, i.e., it acts 
like a correction to the scaling dimension of the system. In fact, if we take 
into account that the canonical dimension of the gauge couplings is given by
\be
d_i:=[\alpha_i] = 4-d,
\ee
then in $d<4$, the beta functions for the gauge couplings feature the leading-order terms
\be
\beta_{\alpha_i} = (d-4)\alpha_i+...
\ee
In $d= 4+\epsilon$, those terms can balance against the terms from quantum 
fluctuations and induce asymptotic safety in pure Yang-Mills theories  
\cite{Peskin1980,Gies:2003ic}. Here, we discover that the quantum-gravity 
correction takes the same form, at least for positive Newton coupling, as a 
\emph{dimensional reduction}. It is intriguing to observe that other forms of 
dimensional reduction have been observed in a variety of quantum-gravity 
approaches in the deep quantum-gravity regime, see, e.g., \cite{Carlip:2016qrb,Ambjorn:2005db,Lauscher:2005qz,Horava:2009if,Calcagni:2013vsa,Carlip:2015mra}.
At this 
point, it is unclear whether a connection between these two observations can be 
established. 
We will now explore the consequences of the effective ``dimensional reduction" encoded in Eqs.~\eqref{eq:betaalphaN} and \eqref{eq:betaalpha2}.

\begin{figure}[t!]
\begin{center}
\includegraphics[width=0.4\linewidth]{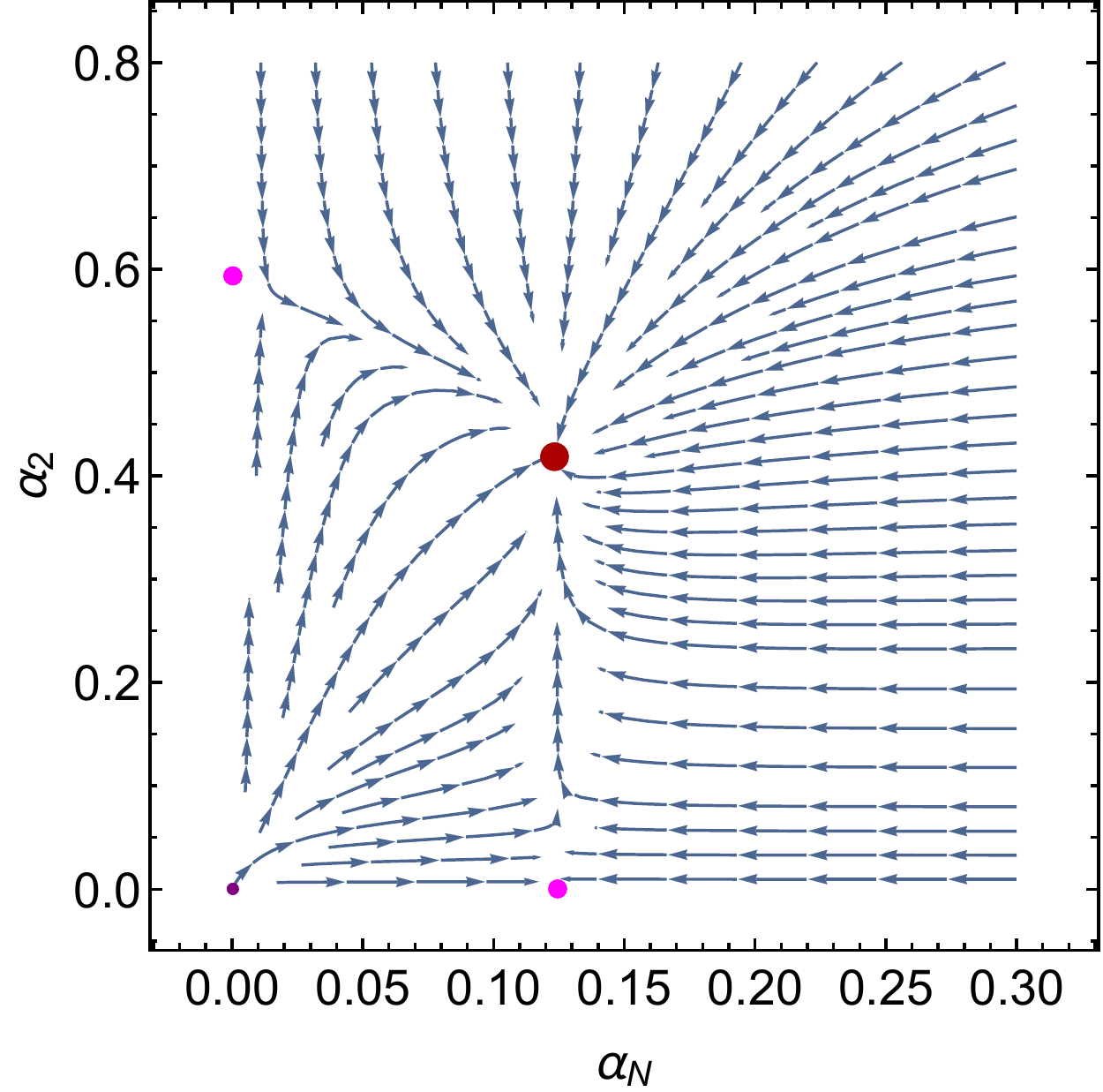}\quad\includegraphics[width=0.4\linewidth]{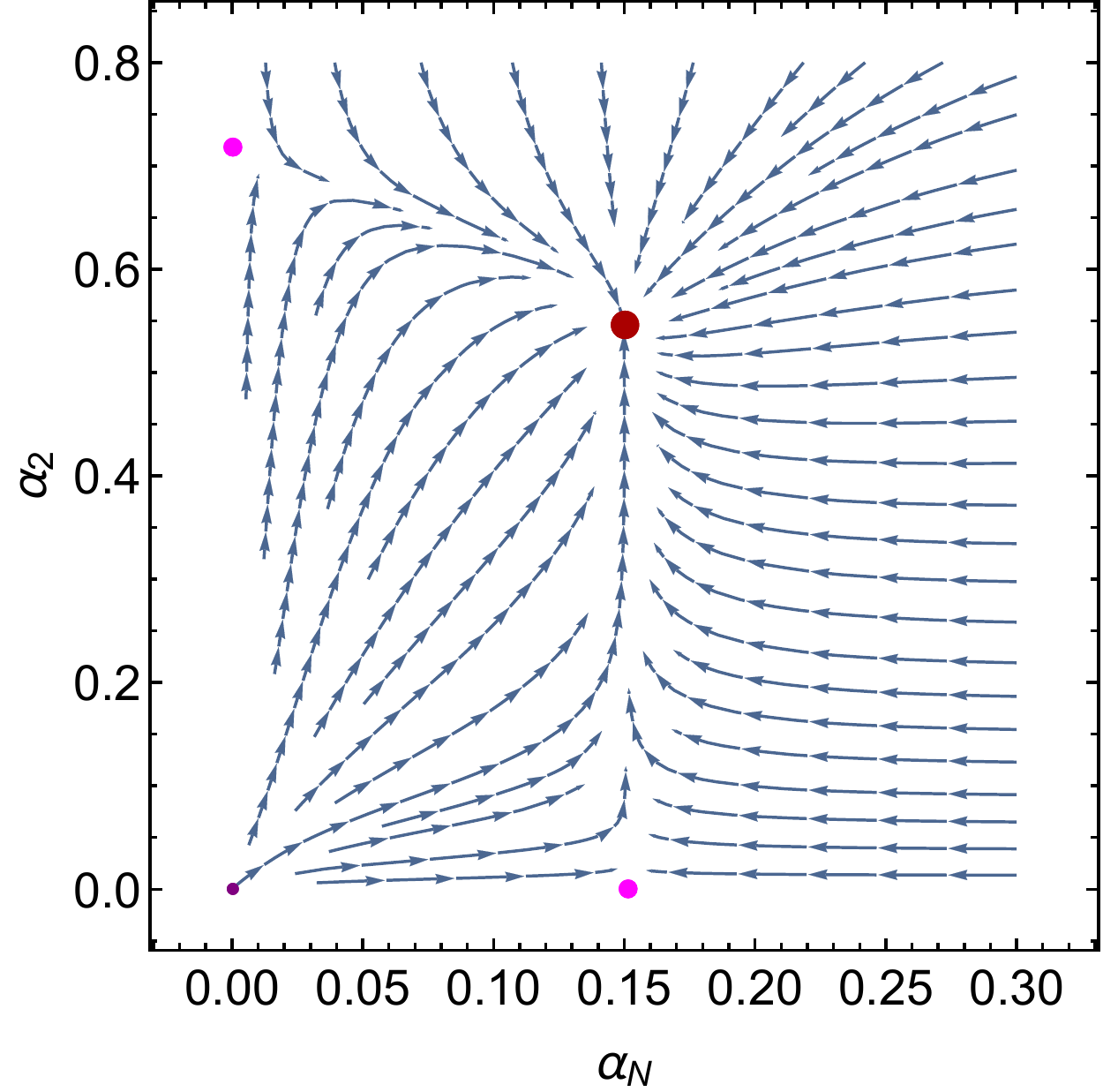}\\
\includegraphics[width=0.4\linewidth]{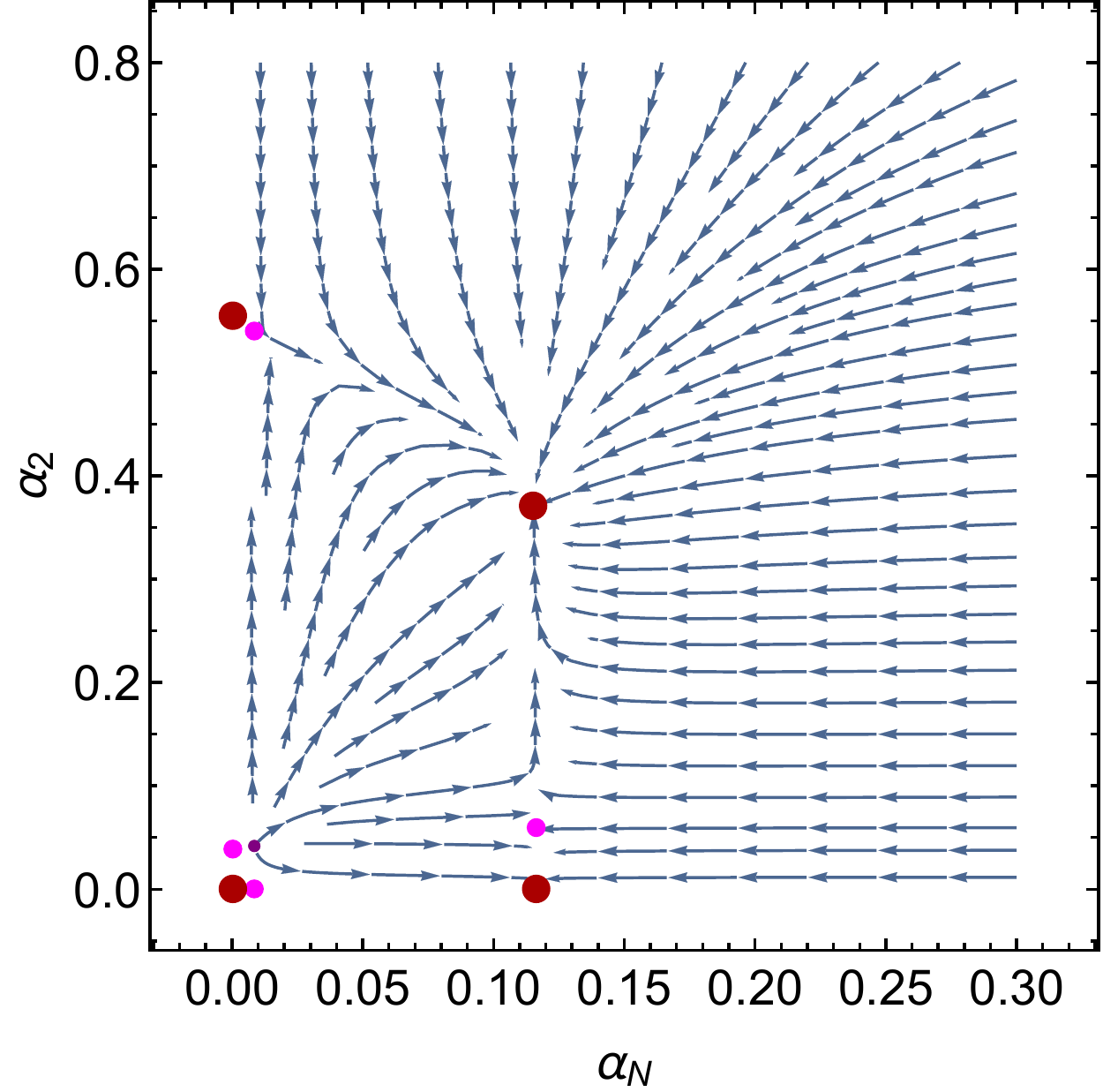}\quad\includegraphics[width=0.4\linewidth]{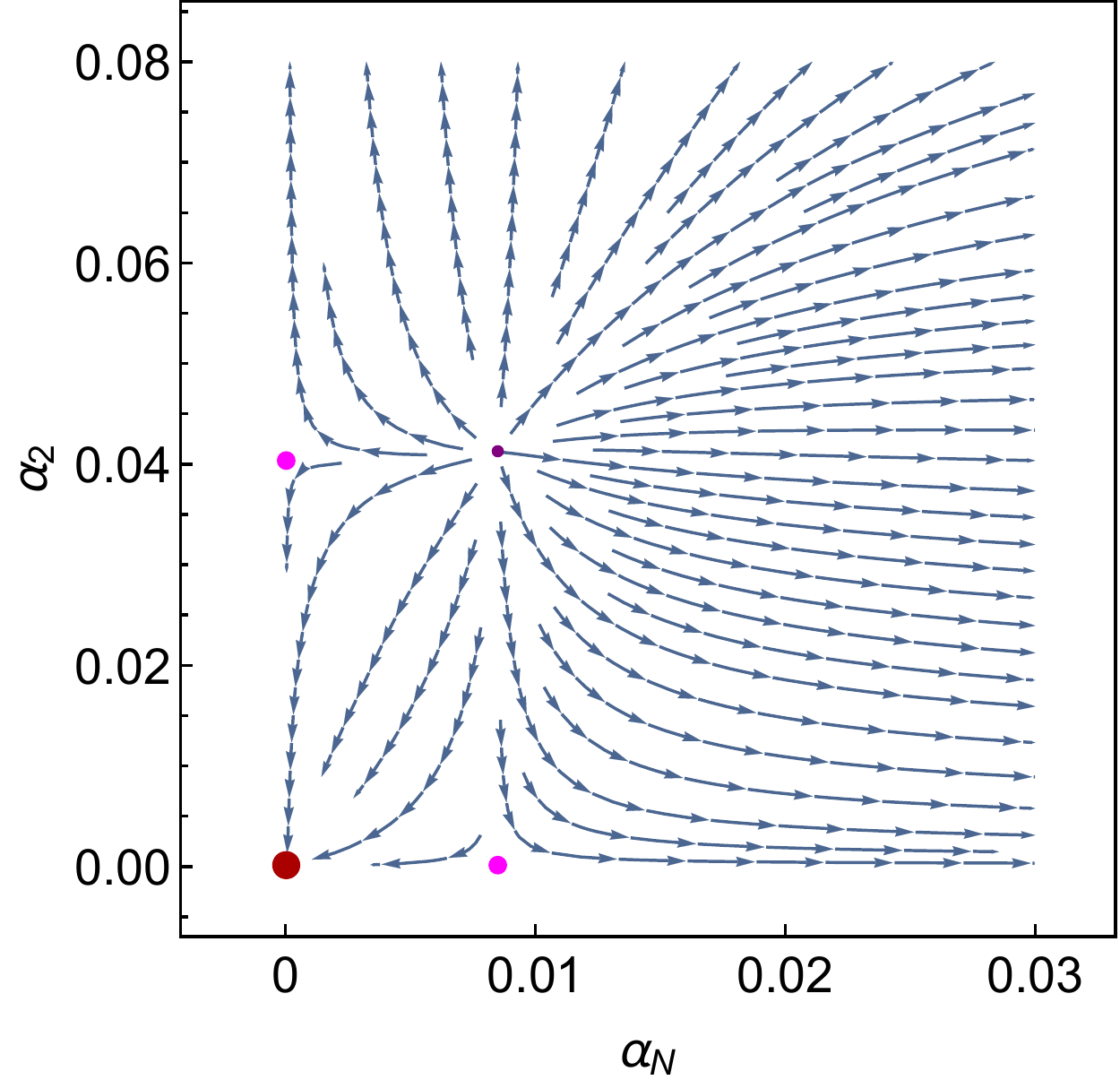}
\caption{\label{fig:flowplots} We show the flow towards the 
IR in the plane spanned by the two gauge couplings for $g=0$ (upper left panel), $g=0.2$ (upper right panel) and $g=-0.05$ (lower left panel)  and 
a zoom in the vicinity of the free fixed point for $g=-0.05$ (lower right panel) for $N=9$ and $M=40$. IR attractive fixed points are shown in dark red (large dot), UV attractive ones in purple (small dot), and mixed (with one UV attractive and IR relevant one) in magenta (medium-size dots).
}
\end{center}
\end{figure}

At the free, i.e., Gau\ss{}ian fixed point, the gravity contribution becomes the leading one. Accordingly, it can have a severe impact on the matter system already at small values of $g$. We observe that the effective dimensional reduction stabilizes the fixed-point structure of the system, cf.~Fig.~\ref{fig:flowplots}: For values of $N, M$ where the free, semi-interacting and interacting fixed points of the system exist without gravity, they also persist at finite $g$ and simply move a little further apart from each other. In particular, as quantum gravity acts in the direction of strengthening asymptotic freedom, the UV/IR-attractivity properties of the various fixed points remain intact: The free fixed point is fully UV attractive, and the interacting fixed point is fully IR attractive, with the semi-interacting fixed points featuring one UV attractive and one IR attractive direction.  Note that under the inclusion of gravity, the fixed points become \emph{partial} fixed points of the full system: As $g$ does not approach a fixed point, but scales canonically, the fixed points in the matter system become scale-dependent. Thus fixed-point scaling in the matter system is no longer signalled by constant dimensionless couplings, but instead by couplings which follow the gravitational scaling through the scale-dependence of the $g$-dependent fixed points.  This ``running scale-invariance" holds at values of $g$ which are not too small. Towards the IR, the standard fixed-point scaling of the pure matter system is restored quickly.

Crucially, the effect of gravity on the system would be completely different if the gravity contribution had the opposite sign -- i.e., if gravity was repulsive instead of attractive -- which we highlight in the lower two panels of Fig.~\ref{fig:flowplots}: The opposite sign leads to a split of the free fixed point into four separate fixed points, one of which is now fully UV attractive. 
In particular, if the gravity correction came with a positive sign, it would 
push the four fixed points towards each other, and induce a fixed-point 
annihilation already at small $g$, leaving behind an IR-attractive 
Gau\ss{}ian fixed point. It is intriguing that the sign of the gravity 
contribution that preserves the fixed-point structure in the system is also the 
one that could induce asymptotic freedom even in Abelian gauge theories 
\cite{Harst:2011zx,Christiansen:2017gtg}.

\begin{figure}[t!]
\begin{center}
\includegraphics[width=0.4\linewidth]{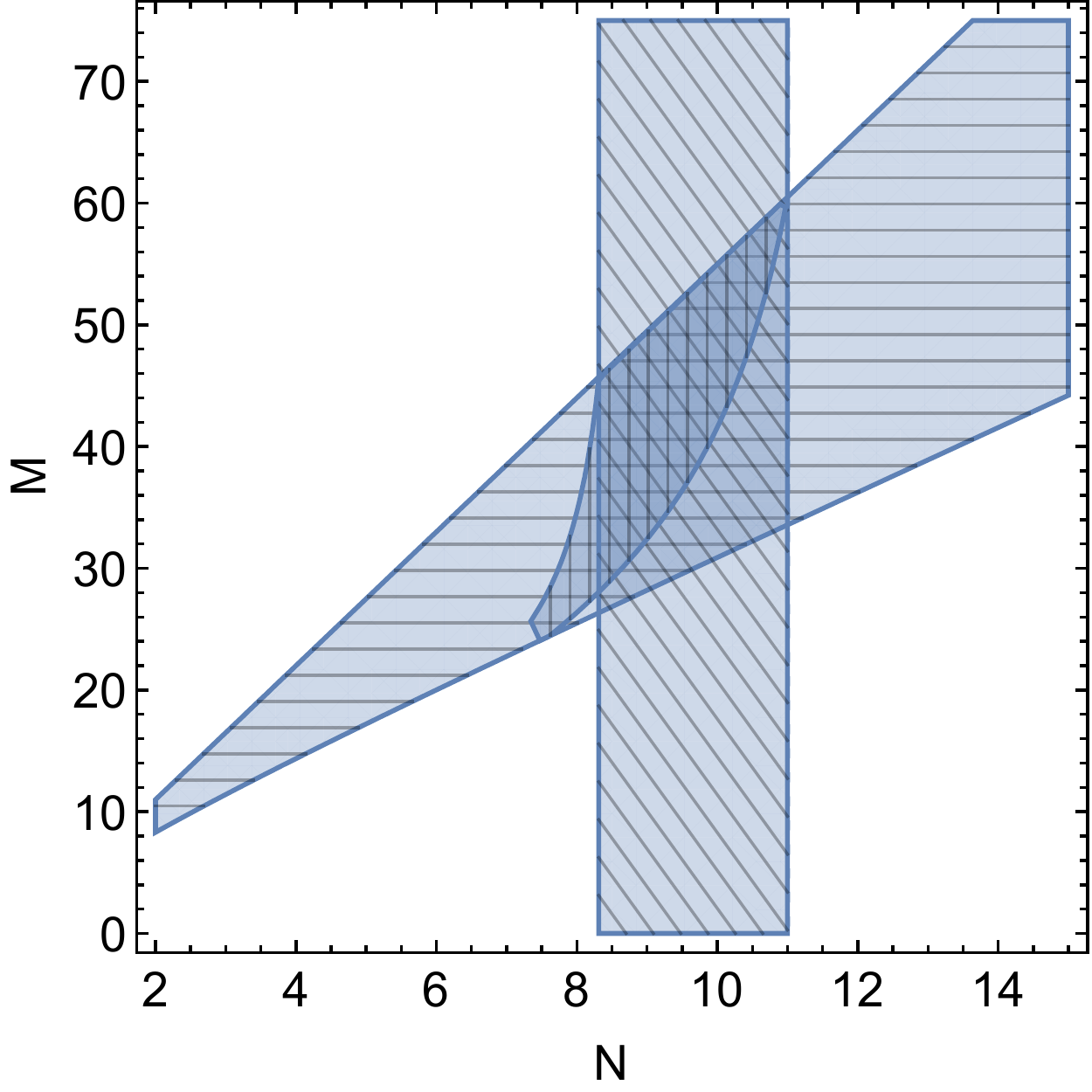}\quad
\includegraphics[width=0.4\linewidth]{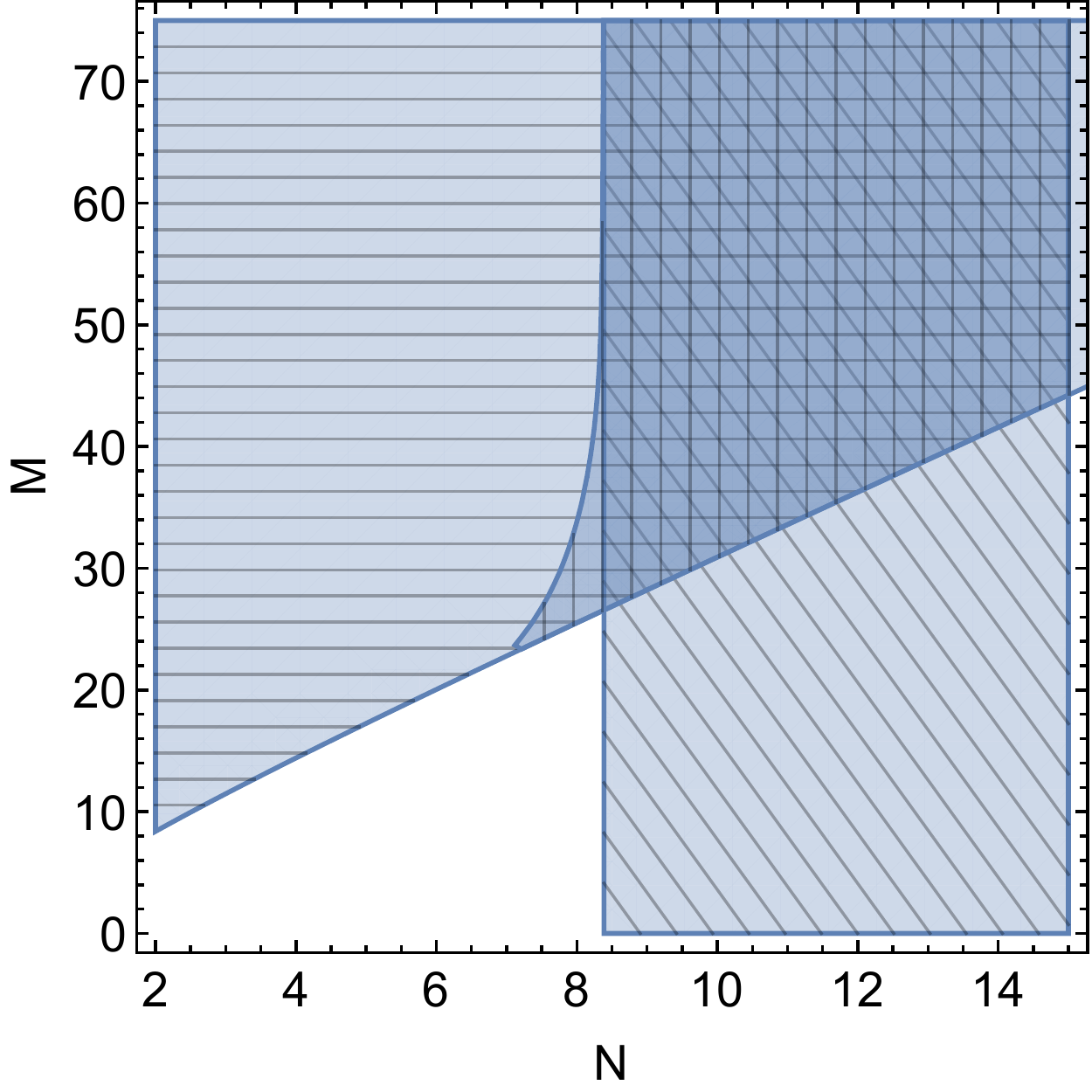}
\caption{\label{fig:FPexistence} We contrast the existence of fixed points in the case without gravity (left panel) to the case with gravity (right panel, $g=0.1$). The existence region of FP1 is shown with horizontal lines, that of FP2 with diagonal lines, and that of FP3 with vertical lines.}
\end{center}
\end{figure}

\begin{figure}[!t]
\includegraphics[width=0.45\linewidth]{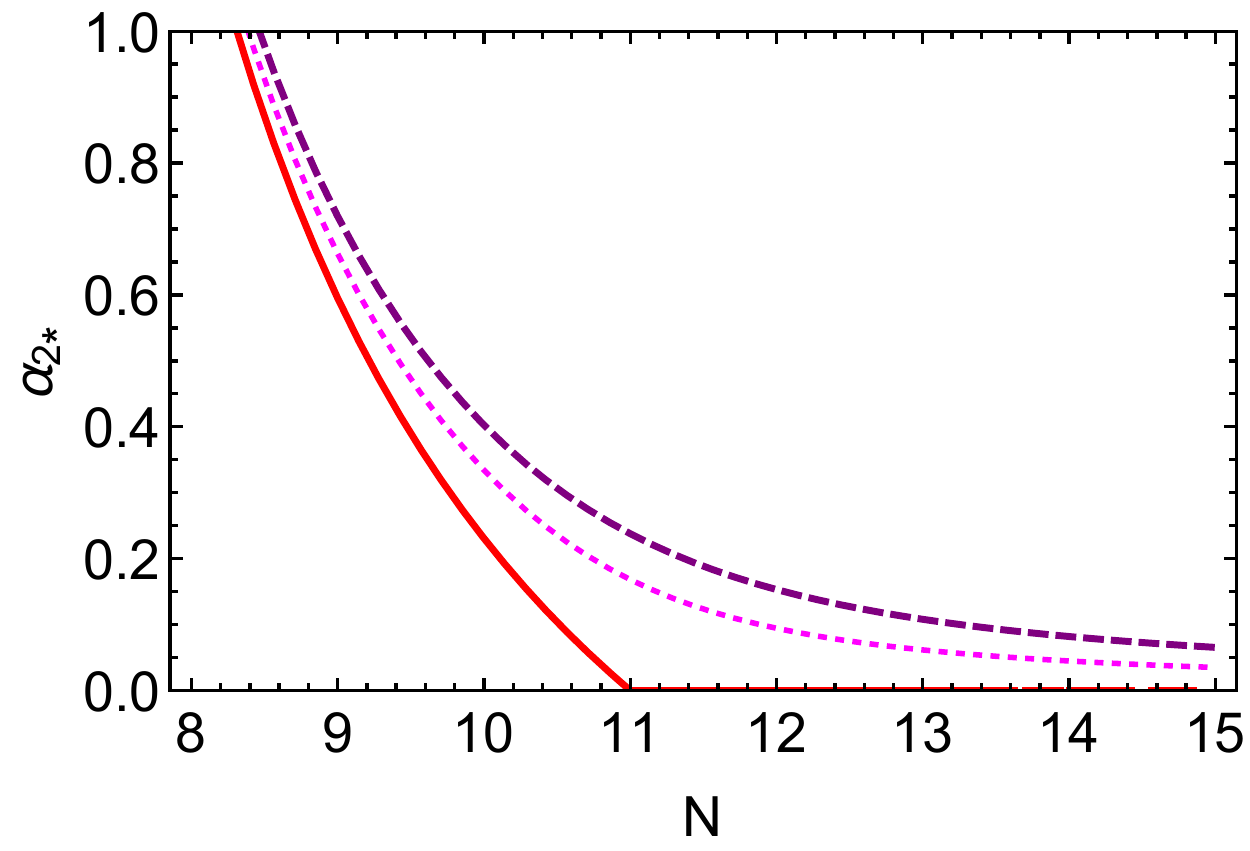}\quad \includegraphics[width=0.45\linewidth]{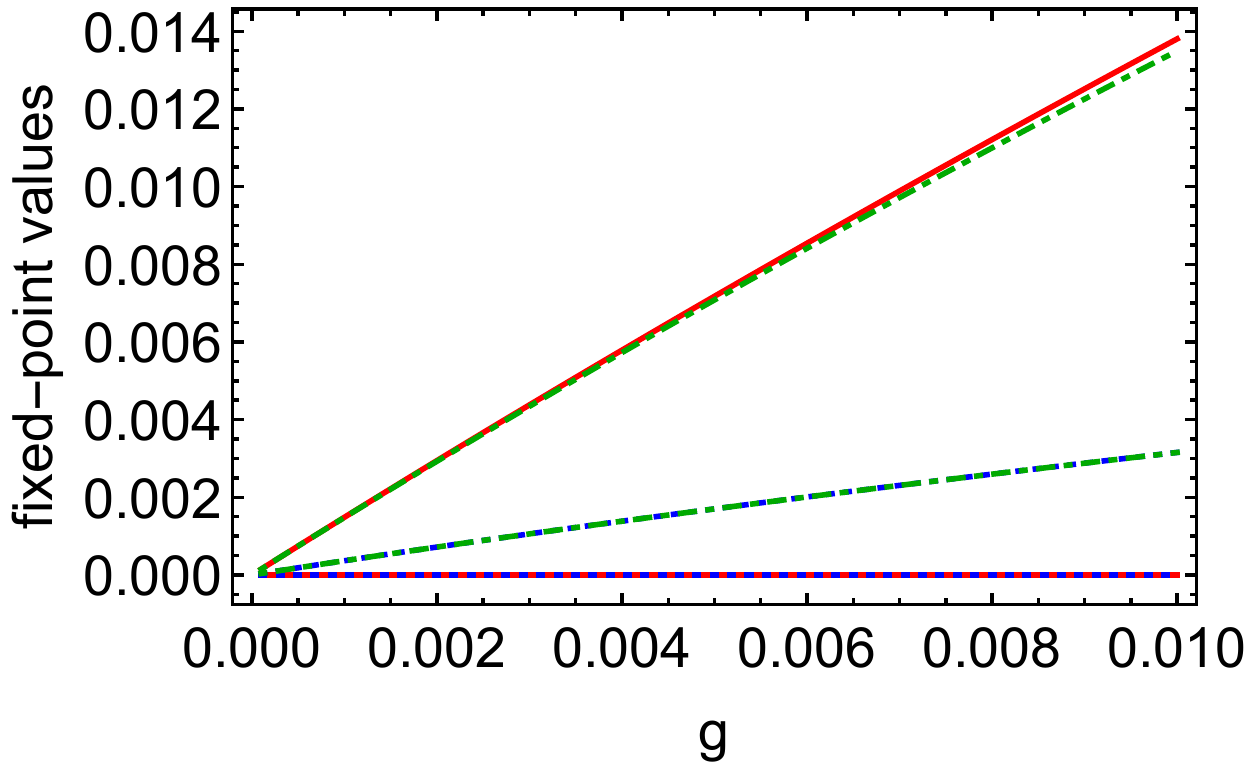}
\caption{\label{fig:semisimpleFPg} Left panel: We show the fixed-point value for $\alpha_2$ at FP1 as a function of $N$ for $g=0$ (thick red line), $g=0.1$ (magenta dotted line) and $g=0.2$ (purple dashed line). Right panel: We show the fixed-point values at FP1 (blue, dotted lines, $\alpha_{2\, \ast2}=0$), FP2 (continuous red lines, $\alpha_{N\, \ast2}=0$), and FP3 (green, dot-dashed lines, $\alpha_{N\, \ast3}<\alpha_{2\, \ast2}$) as a function of $g$ at $N=12,\, M=70$. For $g=0$, all three fixed points lie on top of the Gau\ss{}ian fixed point, whereas they continue to exist as (partially) interacting fixed points if gravity is included.}
\end{figure}

In particular, it seems that quantum gravity effects extend the regions where the various interacting fixed points exist, cf.~Fig.~\ref{fig:FPexistence} and Fig.~\ref{fig:semisimpleFPg}.
 Following the terminology in \cite{Esbensen:2015cjw}, we denote the two fixed points which are UV attractive in one direction by FP1 and FP2, and the fully UV attractive one by FP3. 

 The mechanism is as follows: For instance in the case of FP1, the upper limit on $N$ for which the fixed point is viable at $g=0$ follows from the condition $\alpha_{2\,\ast}>0$. The fixed-point value for $\alpha_2$ decreases as a function of $N$, until FP1 collides with the Gau\ss{}ian fixed point, and tunnels through it to lie at negative values of $\alpha_2$. Intriguingly, the presence of metric fluctuations prevents this fixed-point collision. 
 Hence,
 the fixed point continues to exist for large $N$ and only reaches the Gau\ss{}ian fixed point in the limit $N \rightarrow \infty$.
 To understand the existence of fixed points in a larger region in $N,M$, let us focus on a fixed point which  features $\alpha_{N\, \ast}=0,\alpha_{2\, \ast}>0$.
In that case, it suffices to analyze the effect of quantum gravity on $\beta_{\alpha_2}$, which for $\alpha_N=0$ has the following fixed-point solutions:
\bea
\alpha_{2\, \ast1}&=&0, \quad\alpha_{2\, \ast2/3}=\frac{\left(88-8N\right)\pi \pm \sqrt{8 \pi \left(3g(49N-272)+8\pi(N-11)^2 \right)}}{49 N-272}.\label{eq:alpha2FP}
\eea
The additional term $\sim g$ lifts the degeneracy between $\alpha_{2\, \ast1}$ and $\alpha_{2\, \ast2}$, that exists for $g=0$. Instead, there is one free fixed point and one interacting fixed point at $\alpha_{2\, \ast2}<0$, which remains unphysical. Moreover, $\alpha_{2\, \ast3}>0$ for all $N>5.55$.
Instead of crossing zero at $N=11$, which happens for $g=0$, the fixed-point value for $\alpha_2$ approaches zero asymptotically from the positive regime. The fixed point therefore exists for all values of $N>5.55$, and is perturbative in the sense of $\alpha_{2\, \ast}<1$ for $N > 8.31 + 0.76 g$.
The attractivity properties of the fixed points 
 remain unaltered, as critical exponents only change signs in fixed-point collision and these are avoided here.\\
 To conclude this subsection, we summarize that quantum-gravity effects on the fixed-point structure in semi-simple gauge theories are similar to a (scale-dependent) effective dimensional reduction. They preserve asymptotic freedom, and therefore enlarge the regions in $N,M$ where one totally 
IR- attractive fixed point and two fixed points with one UV attractive direction each exist. 

\section{Quantum-gravity effects on gauge-Yukawa systems}\label{sec:QGgaugeYukawa}
\subsection{Asymptotic safety for gauge-Yukawa systems with gravity}
Following \cite{Litim:2014uca}, we consider a gauge-Yukawa system with Yukawa coupling $y$ and gauge coupling $g$. We define 
\be
\alpha_y = \frac{y^2N_C}{(4\pi)^2}, \quad \alpha_g = \frac{g^2N_C}{(4\pi)^2} 
\,. \label{eq:def_couplings}
\ee
The system contains an SU($N_C$) gauge field, $N_F$ flavors of fermions in the fundamental representation of the gauge group, and $N_F\times N_F$ complex scalar fields which are uncharged under the gauge group.

In \cite{Litim:2014uca}, the $\beta$-functions of this system 
are explored in detail. Here we add the gravity contributions and arrive at
\bea
\beta_{\alpha_N}&=&\alpha_N^2 \Bigl(\frac{4}{3}\epsilon + 
\left(25+\frac{26}{3}\epsilon \right)\alpha_N -2 \left(\frac{11}{2} + \epsilon 
\right)^2 \alpha_y \Bigr) + \beta_{{\alpha_N}}\Big|_{\rm 
grav}\label{eq:betaN}\\
\beta_{\alpha_y}&=& \alpha_y \Bigl(\left(13+2\epsilon \right)\alpha_y - 6 
\alpha_N \Bigr) + \beta_{\alpha_y}\Big|_{\rm grav}.\label{eq:betay}
\eea
The parameter $\epsilon$ is defined as
\be
\epsilon := \frac{N_F}{N_C} - \frac{11}{2} \,.\label{eq:def_epsilon}
\ee
The Veneziano limit \cite{Veneziano:1979ec} is a particular case of a large N 
limit, where the number of colors and flavors is taken to infinity, $N_F 
\rightarrow \infty, \, N_C \rightarrow \infty$, while their ratio is kept 
fixed. 
Hence, the parameter $\epsilon$ can be tuned continuously to any number, in 
particular it can be chosen to be arbitrarily small. This allowed the authors 
of \cite{Litim:2014uca} to use 
$\epsilon$ as a perturbative expansion parameter and opens up the possibility 
to investigate interacting fixed points which are under control in 
perturbation theory.
The quantum-gravity contribution $\beta_{{\alpha_N}}\Big|_{\rm 
grav}$ to the running of the gauge coupling remains 
the same as in 
the previous section, see Eq.~\eqref{eq:betagravalpha}. 
The direct contribution
to the running of a simple Yukawa coupling has been evaluated in 
\cite{Zanusso:2009bs,Vacca:2010mj,Eichhorn:2016esv,Oda:2015sma,Hamada:2017rvn}, including anomalous 
dimensions in \cite{Held:2017}.
Here, we have also derived the gravity contributions to the 
$\beta$-functions in the gauge 
and matter sector for general gauges (see appendix A), thus generalizing the 
results of \cite{Christiansen:2017gtg} and \cite{Held:2017}. For definiteness, 
we now focus on the gauge 
$\alpha=0, \beta=1$, and highlight  in App.~\ref{sec:gaugedependence} that 
none 
of the qualitative features of our 
analysis depends on the choice of gauge. 
The gravity contribution to the running of the Yukawa coupling takes the form
\be
\beta_{\alpha_y}\Big|_{\rm grav} = g\, \alpha_y\, \frac{17}{10\pi}.
\ee
It is  essential
to note that the gravity contributions to the running of 
the couplings in the gauge- and matter sector are \emph{linear} in the gauge 
and matter couplings respectively.  Within our assumption, where the running of $g$ is given by dimensional scaling only,  this implies that the gravity contributions 
cannot be neglected in the large-N limit, e.g., in the Veneziano limit.
In that limit, a simple counting argument might suggest that quantum gravity 
effects should be negligible, as there are $N_F\, N_C$ fermions, $N_C^2-1$ 
gauge fields and $N_F^2$ scalars, whereas there is only one spin-2-mode. 
However, as the quantum-gravity correction is linear in the coupling, it 
acts like a change in the canonical scaling dimension. As the 
canonical scaling dimension of a coupling is of course not affected by the 
number of fields in the system, the quantum-gravity correction is \emph{not} 
negligible in the Veneziano limit. One can also convince oneself of this 
property by starting with the system of couplings $\tilde{\alpha}_g = 
\alpha_g/N_C$, $\tilde{\alpha}_y = \alpha_y/N_C$ and performing the 
redefinition $\tilde{\alpha}_i \rightarrow \alpha_i$, thus making 
the transition to the beta functions in Eqs.~\eqref{eq:betaN}, 
\eqref{eq:betay}. This leads to a suppression of all terms in the beta function 
which do not scale with 
an appropriate positive power of $N_C$, \emph{except} the terms linear in the 
couplings, as in this case the  factors of $N_C$ on the left- and 
on the right-hand side of the $\beta$-functions cancel.

\begin{figure}[!t]
\begin{center}
\includegraphics[width=0.45\linewidth]{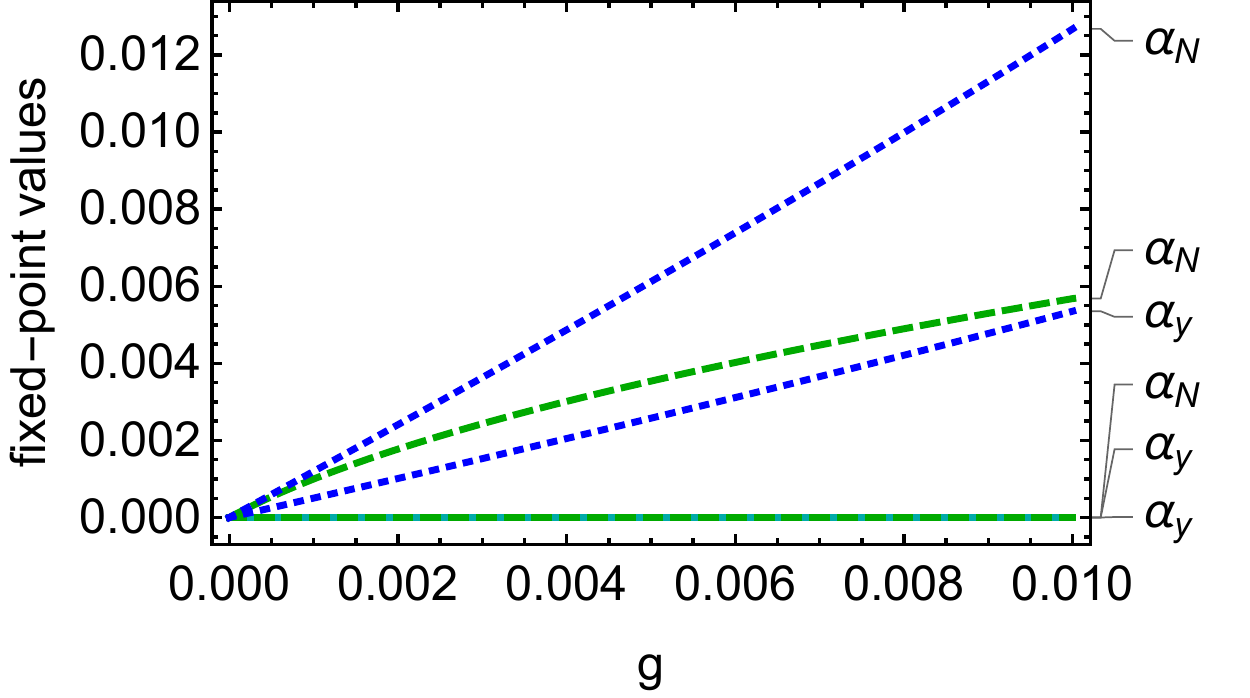}\quad \includegraphics[width=0.41\linewidth]{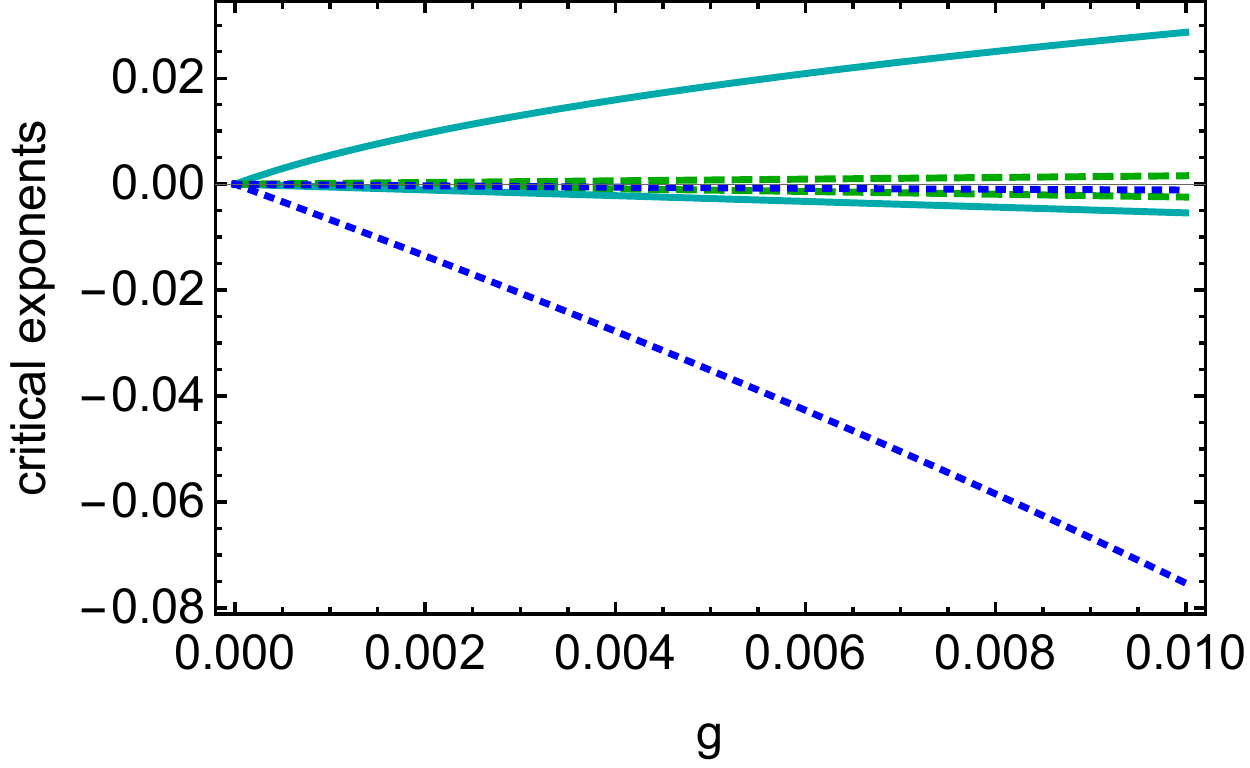}
\end{center}
\caption{\label{fig:FPsplit_beta1_alpha0} We show the split of the Gau\ss{}ian fixed point for $\epsilon=0.1$ as a function of $g$ (fixed-point values for couplings in the left panel and for critical exponents in the right panel). Dotted blue lines correspond to the  IR-attractive interacting fixed point in Eq.~\eqref{eq:alphaNast} and Eq.~\eqref{eq:alphayast}.}
\end{figure}

\begin{figure}[!t]
\includegraphics[width=0.45\linewidth]{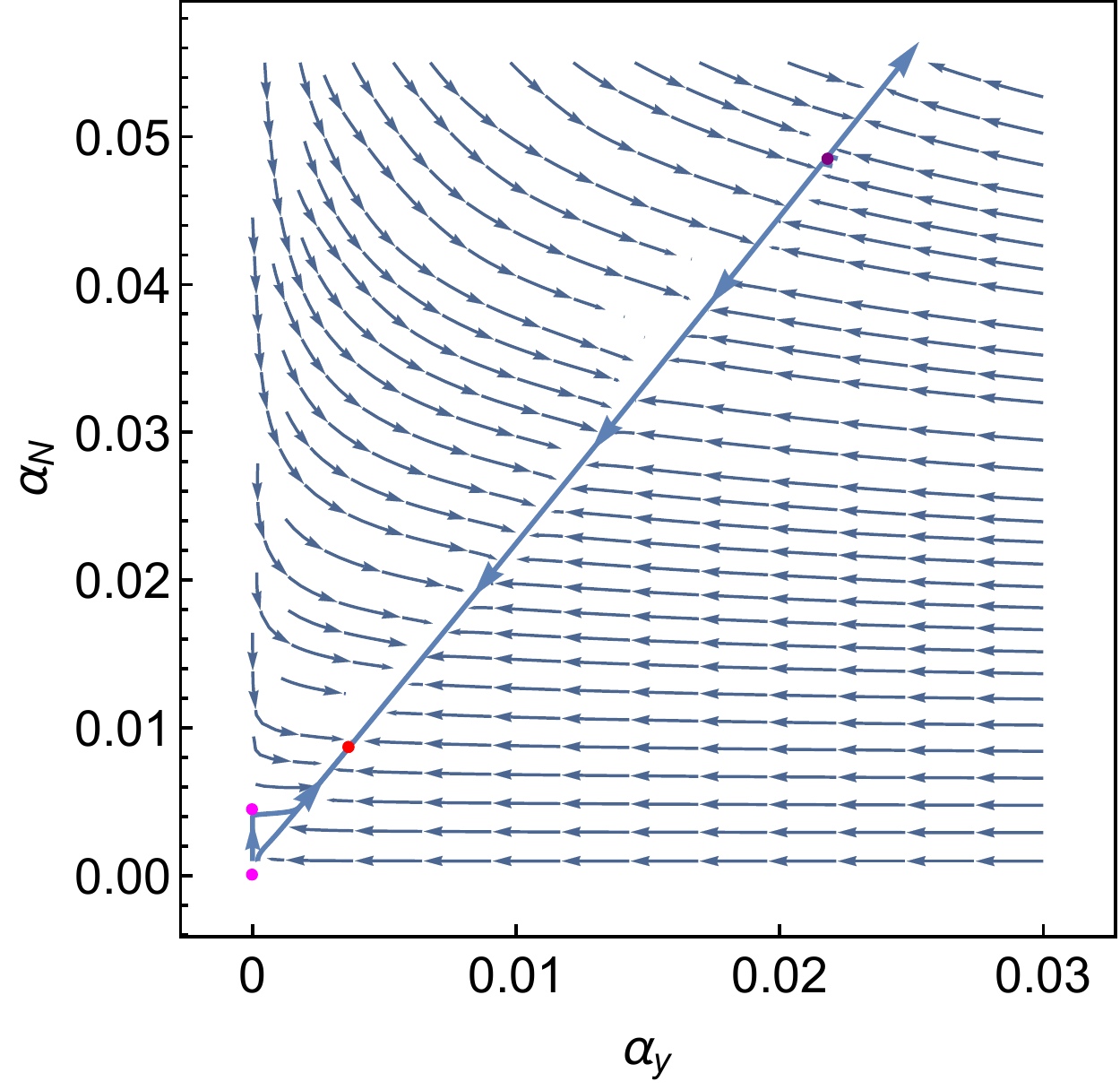}\quad \includegraphics[width=0.45\linewidth]{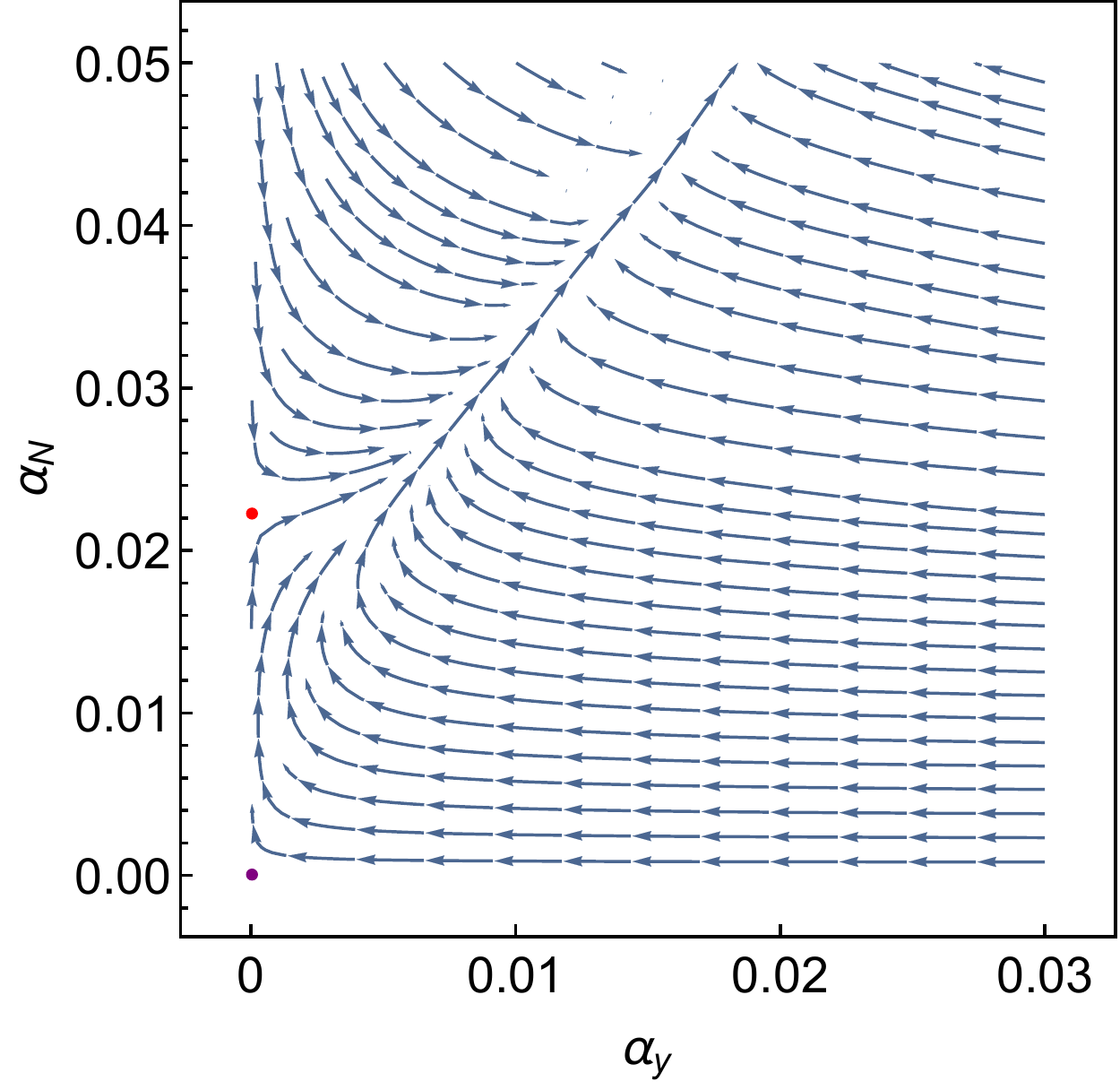}
\caption{\label{fig:flowplot} We show the flow in the plane of the two couplings for $\epsilon=0.1$ and fixed $g=7\cdot 10^{-3}$ (left panel), where a fully interacting IR attractive fixed point exists in addition to the fully interacting UV attractive one; and the case with $g=0.1$ (right panel), where the two fully interacting fixed points have annihilated, leaving behind a free fixed point and a partially interacting one, both of which have one UV attractive direction each.}
\end{figure}

 We observe that
the gravity contributions to different matter couplings differ in their sign, 
thus the effective  scaling dimensionality  that combines the canonical dimension with the quantum-gravity induced one is different for each operator: While 
quantum-gravity corrections to the scale dependence of the Yukawa coupling act 
as if the effective dimension is increased, those to the scale dependence of the 
gauge coupling again act like a dimensional reduction \cite{Codello:2016muj}. Note that it is this 
effective dimensional reduction in the scaling dimension of the gauge coupling 
which fundamentally alters the phase structure of the gauge-gravity-system, even 
for small values of the Newton coupling: As gravity acts to strengthen 
asymptotic freedom, which stabilized the gauge system in the previous section, 
gravity is responsible for a \emph{destabilizing} effect in the gauge-Yukawa 
system:
In the presence of the linear gravity term, the degeneracy of the free fixed point is lifted and it splits into new interacting fixed points. These emerge in the regime $\alpha_N>0, \alpha_y>0$, cf.~Fig.~\ref{fig:FPsplit_beta1_alpha0} and \ref{fig:flowplot}. 
Specifically, the coordinates of a new, fully interacting fixed point are 
\bea
\alpha_{N\, \ast}&=&\frac{1}{80\pi(57-46\epsilon - 8\epsilon^2)}\Bigl[ 51 g (11+2\epsilon)^2 + 80 \pi \epsilon (13+2\epsilon) \label{eq:alphaNast}\\
&{}&- \sqrt{4800\pi g(13+2\epsilon)(-57+46\epsilon+8\epsilon^2) 
+(51g(11+2\epsilon)^2+80 \pi \epsilon(13+2 \epsilon))^2} \Bigr], \nonumber\\
\alpha_{y\, \ast}&=&\frac{1}{40\pi(13+2\epsilon) (57-46\epsilon- 8 \epsilon^2)} \Bigl[ 240\pi\epsilon(13+2\epsilon)+17 g(861+580 \epsilon + 68\epsilon^2)\label{eq:alphayast}\\
&{}& -3 \sqrt{4800\pi g(13+2\epsilon)(-57+46\epsilon+8 \epsilon^2)+(51g(11+2\epsilon)^2+80 \pi \epsilon(13+2 \epsilon))^2}\nonumber
\Bigr],
\eea
which clearly approaches the free fixed point in the limit $g\rightarrow 0$. We do not explicitly provide the expression for the critical exponents, which is rather lengthy, however they are both negative, when the fixed point is real. The critical exponents are defined via the stability matrix
\bea
\theta_{1,2} = - {\rm eig} \left( \begin{array}{cc}
\frac{\partial \beta_{\alpha_N}}{\partial \alpha_N} & \frac{\partial \beta_{\alpha_N}}{\partial \alpha_y}\\
\frac{\partial \beta_{\alpha_y}}{\partial \alpha_N} & \frac{\partial \beta_{\alpha_y}}{\partial \alpha_y}
\end{array}
\right)\Bigg|_{\alpha_N=\alpha_{N\, \ast},\, \alpha_y = \alpha_{y\,\ast}},
\eea
such that a negative critical exponent corresponds to an IR attractive direction.

If the sign of the gravity-contribution to the gauge coupling was opposite, all the interacting fixed points which emerge from the free one under the impact of gravity would lie in the physically unacceptable regime $\alpha_{N\,\ast}<0$, as they do in the case of semi-simple gauge theories in the previous section. 
 This implies that the dynamics of the system closely resembles that of the system in $d=4-\epsilon$ dimensions analyzed in \cite{Codello:2016muj}.
 
 The existence of the new fixed point can easily be understood from the structure of the beta function in the gauge sector, cf.~Eq.~\eqref{eq:betaN}. Schematically, it is given by
\be
 \beta_{\alpha_N} = -\mathcal{D}^{\mathrm{grav}}_{\alpha_N}g \, \alpha_N- B\, \alpha_N^2 + 
C\,\alpha_N^3,
 \ee
where the coefficient $\mathcal{D}^{\mathrm{grav}}_{\alpha_N}$ parameterizes  
the gravitational contributions to the running of the gauge coupling 
$\alpha_N$.
 An asymptotically safe UV fixed point arises in the case of vanishing gravity, $g=0$, if $B<0$ and $C<0$, which can be arranged with the help of the Yukawa coupling. With gravity, it is instructive to rewrite the beta function in the form
 \be
 \beta_{\alpha_N} =\alpha_N(-\mathcal{D}^{\mathrm{grav}}_{\alpha_N}g -  \alpha_N \left(B- 
C\, \alpha_N \right)).
 \ee
 The fixed point at $\alpha_N=0$ does not change its location, but becomes UV 
attractive for $\mathcal{D}^{\mathrm{grav}}_{\alpha_N}g>0$. The  interacting UV 
fixed point at $\alpha_N = \frac{B}{C}$ is shifted and lies at
 \be
 \alpha_{N\ast\, 1}=\frac{B}{2C} + \sqrt{\frac{B^2}{4 C^2} + 
\frac{\mathcal{D}^{\mathrm{grav}}_{\alpha_N}g}{C}}.
 \ee
For small $g$, the shift is very mild, and the fixed point still lies in the vicinity of its previous value. Further, it remains UV-attractive, as gravity does not change the slope of the beta-function for large $\alpha_N$, since the leading large-$\alpha_N$ term is $g$ independent, and thus the outermost fixed point remains UV attractive. Since we now have two fixed points which are both UV attractive, it is obvious that there must be a third fixed-point solution which must lie inbetween those two. In fact, it is of course given by
\be
\alpha_{N\ast\, 2}= \frac{B}{2C}-\sqrt{\frac{B^2}{4 C^2} + 
\frac{\mathcal{D}^{\mathrm{grav}}_{\alpha_N}g}{C}},
\ee
which clearly emerges from the free fixed point as $g$ increases from zero. As $g$ increases, $\alpha_{N\ast\,2}$ approaches $\alpha_{N \ast\, 1}$, until they annihilate and move into the complex plane at 
\be
g_{\rm crit} = - \frac{B^2}{4C\, \mathcal{D}^{\mathrm{grav}}_{\alpha_N}},
\ee
which is of course positive for $C<0$ and $\mathcal{D}^{\mathrm{grav}}_{\alpha_N}>0$. This 
analysis suggests that quantum gravity can have a severe impact on the system 
and destroy the UV completion of the pure matter system, substituting if for an 
asymptotically free one, as the asymptotically free fixed point is of course 
left behind when the two interacting fixed points annihiliate.

As we are analyzing the EFT regime for gravity, none of the two UV fixed points can of course be a true UV fixed point of the full system.
Instead, a UV completion for gravity is required beyond the Planck scale. The 
(partial) UV fixed points of the matter system could then arise at intermediate 
scales between the Planck scale and the electroweak scale and dominate the 
matter sector in that regime. 
In a 
particularly intriguing case, the UV completion for gravity might also be 
provided by the asymptotic- safety paradigm.

In the regime of small enough Newton coupling, we thus encounter a scenario where the interacting UV fixed point still exists in all cases. Towards the IR, the RG flow can move along the separatrix, which at finite $g$ connects the interacting UV to an interacting IR fixed point. As the scale decreases, the Newton coupling decreases further, such that the interacting IR fixed point moves back into the free fixed point. Thus, we have a setting where the flow approaches a partial fixed point of the system (it is of course not a fixed point in $g$), and that partial fixed point simultaneously moves away. In the deep IR, gravity fluctuations have of course switched off, and the remaining IR fixed point is a Gau\ss{}ian one.

On the other hand, as $g$ grows, the interacting UV and IR fixed points can approach each other. Beyond a critical $g$, these two annihilate, as is signalled by a vanishing critical exponent, and they move into the complex plane, where they cease to be physically acceptable fixed points.

According to Eq.~\eqref{eq:runningg}, the value of $g$ can be translated into a scale. In units of $M_{\rm Planck}$, $g=1$ exactly at the Planck scale and $g<1$ for  
$k<M_{\mathrm{Planck}}$.
Hence, the critical value of $g_{\rm crit}$, beyond which fixed-point annihilation occurs, can be translated into a scale $k_{\rm crit}$: If the UV safe fixed point in the matter sector is to be phenomenologically relevant, it has to be reached at a scale below $k_{\rm crit}$.

\begin{figure}[!t]
\includegraphics[width=0.45\linewidth]{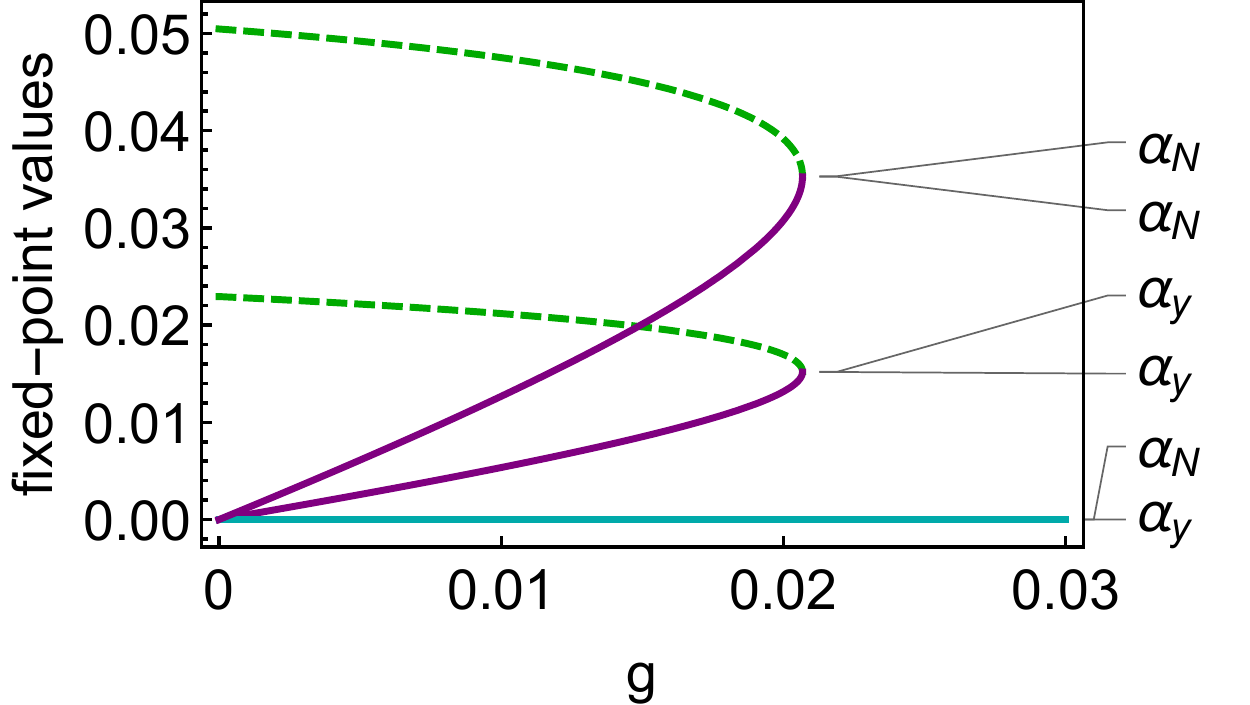} \quad \includegraphics[width=0.45\linewidth]{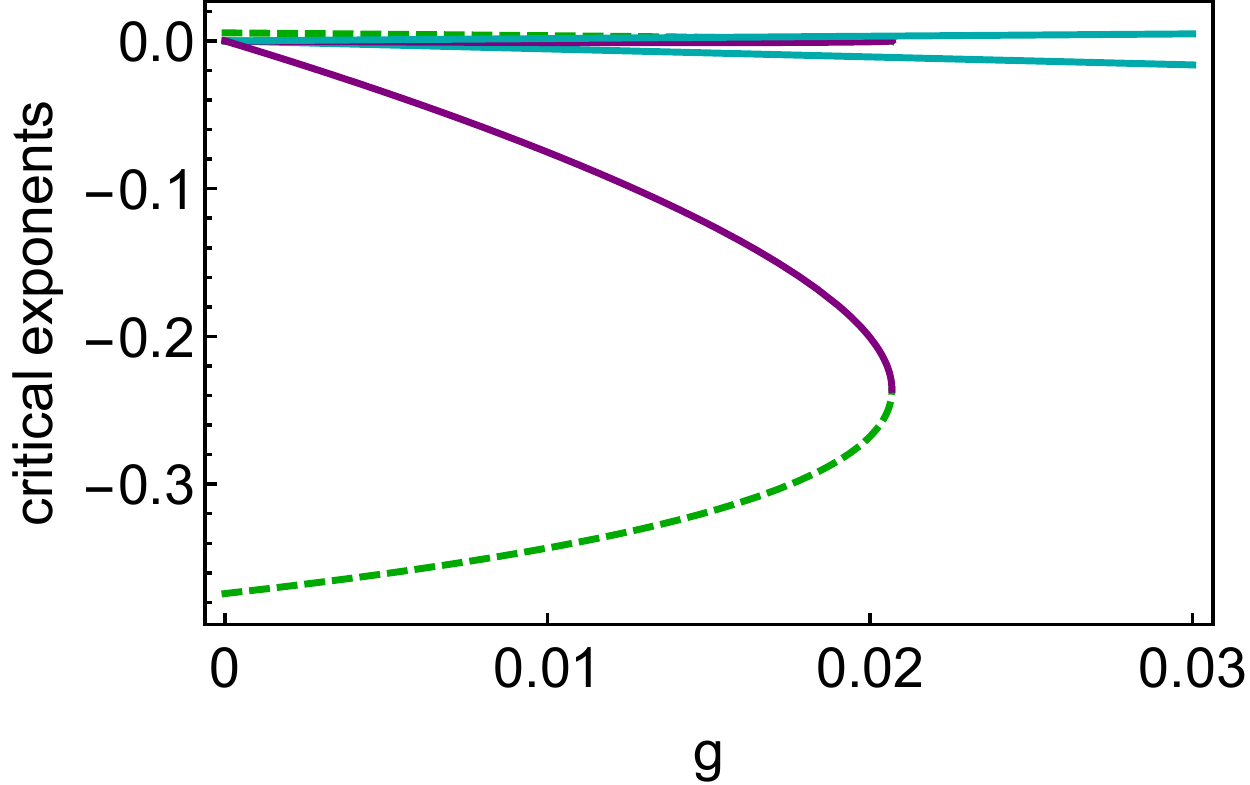}
\caption{\label{fig:FP_anni_beta1_alph0} We show the annihilation between the UV attractive and the IR attractive interacting fixed points at finite $g$ for $\epsilon =0.1$ (left panel). The critical exponents (right panel) also show the same annihilation.}
\end{figure}

The critical value of $g$ at which the fixed-point annihilation occurs is a function of $\epsilon$,
\bea
g_{\rm crit}&=& \frac{80\pi}{867 (11+2\epsilon)^4} \Bigl( 7410 \nonumber\\
&{}&- \sqrt{20(13+2 \epsilon)^2 (-57+46 \epsilon + 8 \epsilon^2)(-285 +\epsilon(2287 +788 \epsilon + 68 \epsilon^2))}\nonumber\\
&{}&- \epsilon \left(31581 + 2 \epsilon (7899 + 2 \epsilon(635 + 34 \epsilon)) \right) \Bigr).\label{eq:gcrit}
\eea
In the limit $\epsilon \rightarrow 0$, where the fixed point 
in the pure matter system moves arbitrarily close to the free fixed point, an 
infinitesimal value of $g$ is sufficient to trigger a fixed-point annihilation,
\begin{equation}
\lim_{\epsilon \rightarrow 0} g_{\mathrm{crit}} =0 \,.
\end{equation}

The fixed points reappear at a larger value of $g_{\rm reappear}$. 
That value
hardly depends on $\epsilon$ and lies at about $g_{\rm reappear} \approx 0.29$, 
until it drops sharply at about $\epsilon \approx 0.1$. For all values of $\epsilon$, the value of 
$g_{\rm reappear}$ lies dangerously close to the Planck scale, where it 
is unclear whether our leading-order treatment of quantum gravity fluctuations is still 
justified. Hence we emphasize that the reappearance of the fixed points might be 
an artifact of the approximation.

\begin{figure}[!t]
\begin{center}
\includegraphics[width=0.45\linewidth]{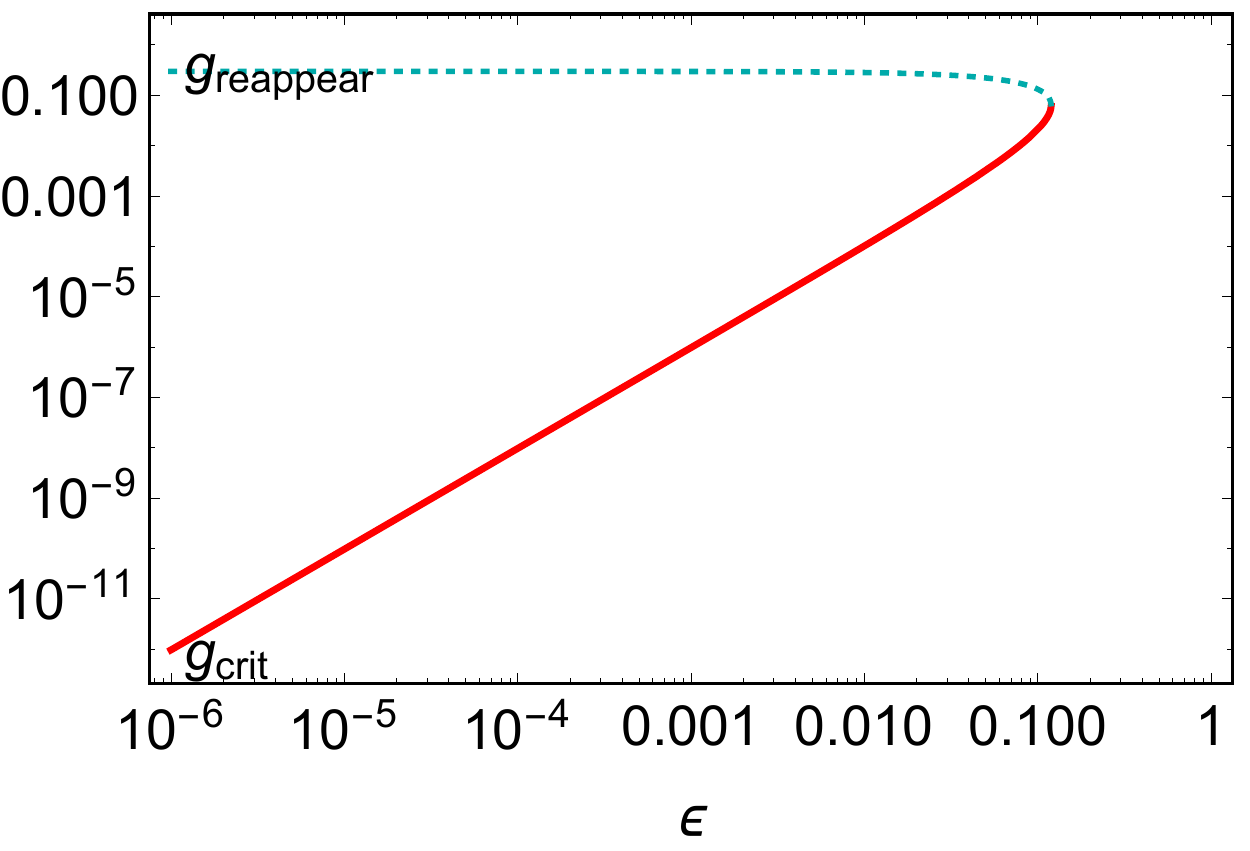}
\end{center}
\caption{\label{fig:gcrit}We show $g_{\rm crit}$ and $g_{\rm reappear}$ as a function of $\epsilon$.}
\end{figure}

\begin{figure}[!t]
\begin{center}
\hspace{-20pt}
\includegraphics[width=0.42\linewidth]{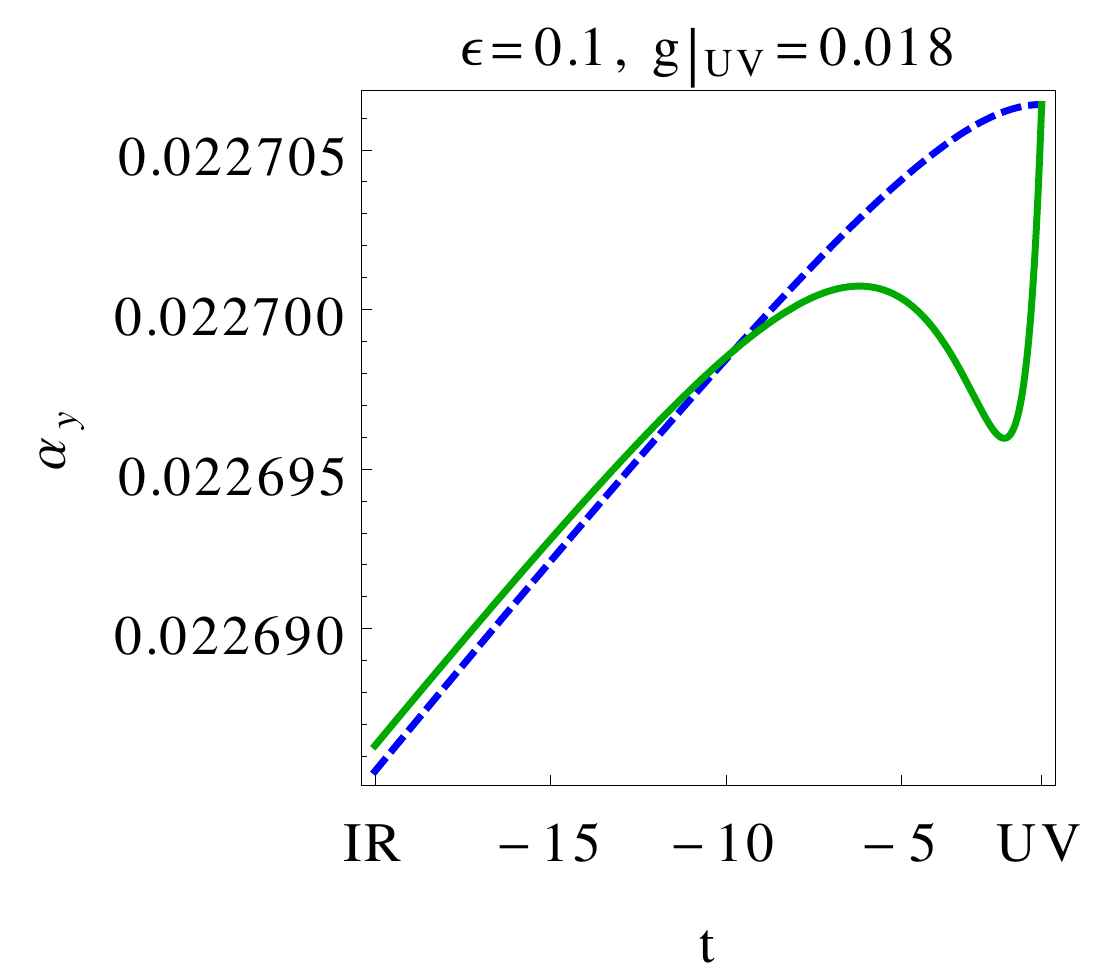}
\hspace{5pt}
\includegraphics[width=0.41\linewidth]{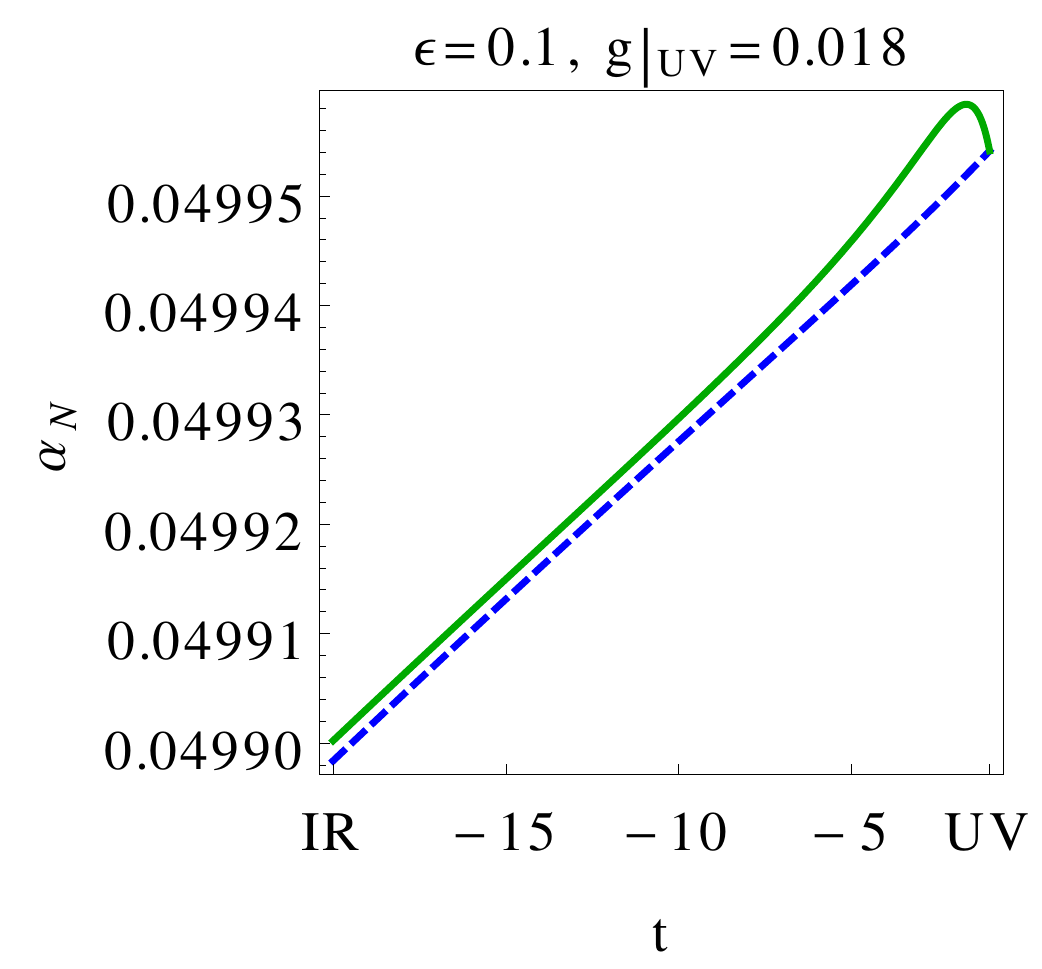}
\end{center}
\caption{\label{fig:trajectories1} We compare the running of $\alpha_N$ and $\alpha_y$ as a function of the RG ``time" $t= \ln(k/k_0)$ for the case with and without gravity. We fix the scales by choosing $g|_\text{UV}=g(k_0)$. Without gravity (blue dashed lines) we initiate the flow close to the UV fixed point. Starting the flow with gravity (green continuous line) at the same initial condition, gravity fluctuations lead to modifications in the flow, before they switch off and the system is dominated by matter-induced running.
}
\end{figure}

\begin{figure}[!t]
\begin{center}
\includegraphics[width=0.37\linewidth]{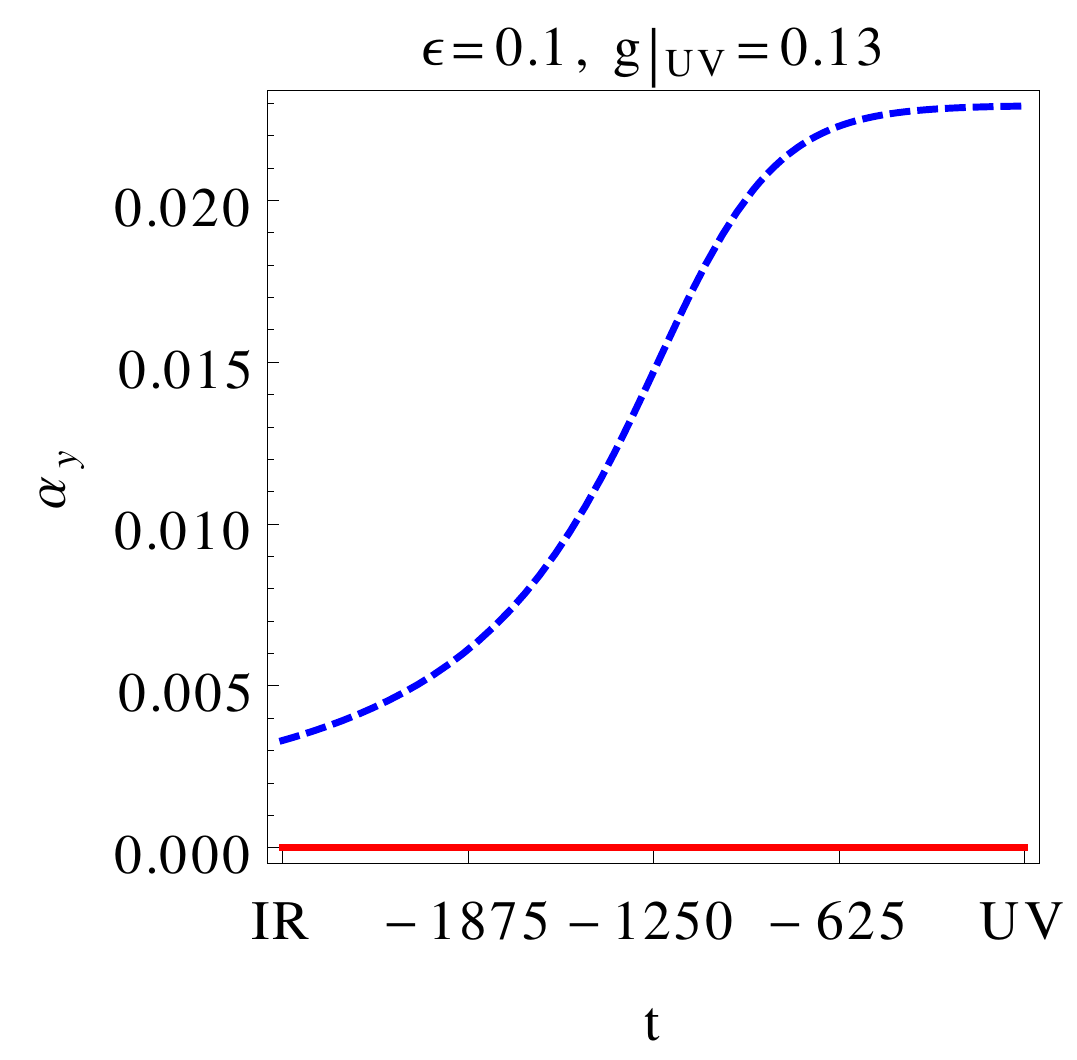}
\hspace{30pt}
\includegraphics[width=0.35\linewidth]{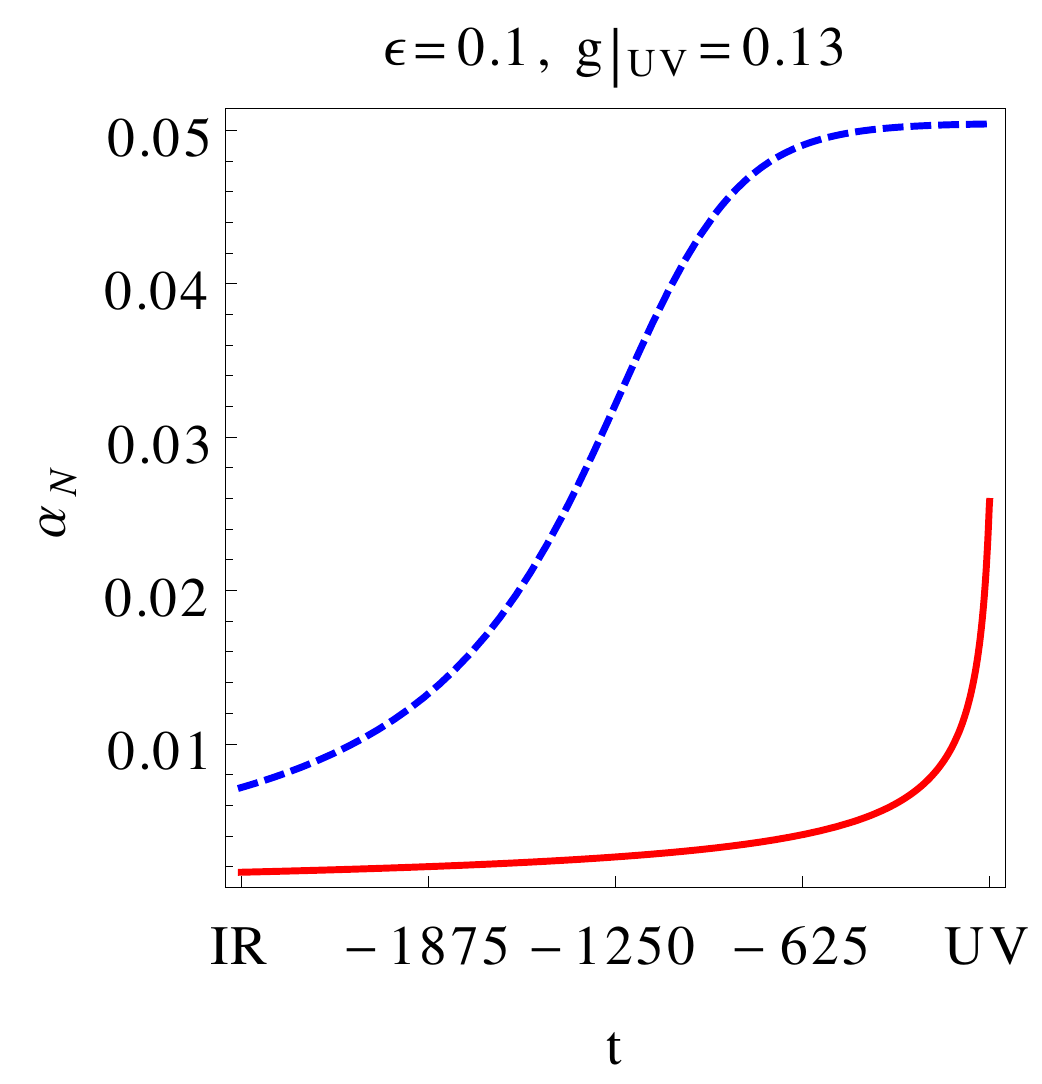}
\end{center}
\caption{
	\label{fig:trajectories}
We compare the running of $\alpha_N$ and $\alpha_y$ as a function of the RG ``time" $t= \ln(k/k_0)$ for the case with and without gravity. We fix the scales by choosing $g|_\text{UV}=g(k_0)$. We initiate the system at different UV starting points, such that the initial condition always corresponds to a UV fixed point. With gravity (red continuous line) and $g>g_{\rm crit}$, the fixed point is the partially interacting one. As a function of the scale, the fixed point moves, since $g$ runs. In the IR, the same universal scaling regime is approached.  Note that the flow towards the IR follows the $g$-dependent fixed point. This induces a much faster running in the case with gravity than the running along the separatrix in the gravity-free case, as $g$ runs quadratically with scale $k$.
}
\end{figure}

As gravitational contributions alter the UV behavior of the matter system, they can prevent the approach to the scale-invariant regime that would be reached in the pure matter system. Instead, a new scaling regime can be induced by quantum gravity fluctuations.  
If we assume a classical regime for gravity, $g$ runs with $k^2$. In that setting, 
the new fixed points inherit the scaling,  exhibiting a form of ``running scale-invariance".  For $g=\rm const$, a standard form of scale invariance is recovered.
Explicit trajectories in the different regimes of gravitational coupling strength ($g=0$, $g<g_\text{crit}$ and $g>g_\text{crit}$) all share the same scale-invariant behavior in the IR, cf.~Fig.~\ref{fig:trajectories1}~\&~\ref{fig:trajectories}. This follows as gravity is dynamically switched off towards the IR,
which is encoded in the canonical running of the Newton coupling $g \sim k^2$, cf. Eq.~\eqref{eq:runningg}.  Here, we focus on that part of the phase diagram in which the matter system becomes weakly coupled in the IR.
In turn, the different types of matter UV-completions that are realized for different strengths of gravitational fluctuations 
 become apparent in signatures of UV 
 fixed-point behavior
 at higher scales. While both, the pure-matter model ($g=0$) and the weakly gravitating model ($g<g_\text{crit}$) exhibit similar fixed-point scaling towards the UV, the stronger gravitating system ($g>g_\text{crit}$) approaches a different UV fixed point (as the original one annihilated, cf.~Figure~\ref{fig:FP_anni_beta1_alph0}). These different UV scale-invariant regimes are 
 distinguishable. Therefore, UV matter-models like the gauge-Yukawa-model at hand in principle offer a
 setting in which
 tests of 
 quantum-gravity effects below the Planck scale could appear feasible. Of course this would require that the corresponding matter models are realized beyond the Standard Model, and moreover this does \emph{not} provide a way to test different UV completions for quantum gravity. 
 Our analysis suggests that gravitational signatures, namely the modified UV-scaling in Fig.~\ref{fig:trajectories1} and \ref{fig:trajectories}, are present at the scale $k_\text{UV-scaling}$ where the matter-model starts to flow towards its UV fixed point. This scale might well be significantly below the Planck scale, i.e., $k_\text{UV-scaling}<M_\text{Planck}$.
 
Finally, let us add the observation that in the original models discussed in \cite{Litim:2014uca}, an Abelian gauge group poses a challenge, as it is not possible to flip the sign of the one-loop coefficient of the beta function by increasing the number of charged fermions -- a prerequisite that is necessary to obtain asymptotic safety in the non-Abelian gauge couplings. Here, we observe that we can generate a partial fixed point for a gauge-Yukawa-system with an additional Abelian gauge group by including quantum-gravity effects. The fixed point will be asymptotically safe in the non-Abelian gauge coupling and the Yukawa coupling, but asymptotically free in the Abelian gauge coupling. As it becomes asymptotically free, the Abelian gauge coupling cannot alter the fixed-point structure in the other two couplings, as it vanishes at the fixed point. Moreover, the gravity-contribution to the beta function of the Abelian gauge coupling is the leading one at the free fixed point, i.e., the contributions from the other gauge coupling and the charged matter fields are completely subleading, and gravity induces asymptotic freedom for arbitrary matter content.

\subsection{Connection to asymptotically safe quantum gravity with matter}
The setting that we have explored so far treats quantum gravity within an effective-field theory regime. Such a treatment is expected to break down at the Planck scale. Beyond, the asymptotic- safety paradigm could provide a UV complete model of quantum gravity and matter. Here, we will briefly highlight how the fixed-point structure in gauge-Yukawa models could match onto a setting with asymptotically safe gravity in the UV. In such a case, the IR could be reached via a cascade of fixed points, starting with a fully asymptotically safe fixed point in the far UV and passing by a fixed point of the gauge-Yukawa system with running $g$.  The more RG ``time" the flow spends in the vicinity of intermediate fixed points, the more one expects universality to set in, i.e., information on the underlying UV theory would be washed out and replaced by information on the universal scaling behavior in the vicinity of the intermediate fixed point.
It is intriguing to observe that the fixed-point structure of the gauge-Yukawa system at $g>g_{\rm crit}$ appears to potentially 
be attainable at intermediate scales if one starts from an asymptotically safe matter-gravity model: There, studies indicate that the Yukawa coupling generically features a fixed point at $\alpha_y=0$ \cite{Zanusso:2009bs,Vacca:2010mj,Eichhorn:2016esv,Oda:2015sma,Hamada:2017rvn,Held:2017}. Moreover, $\alpha_N$ can feature a fixed point at which it is finite and IR-attrative \cite{Harst:2011zx}, at least in the Abelian case.  For the non-Abelian case, one would expect the possibility of an interacting, UV attractive fixed point within a certain part of the gravitational coupling space.
Starting from such a fixed point, the RG flow 
 might
then be led towards the fixed point at $\alpha_N>0, \alpha_y=0$ in the right-hand panel of Fig.~\ref{fig:flowplot}. That fixed point is of course only a partial fixed point of the full system, as the gravitational coupling no longer lives at a fixed point, but scales with $k$ according to the effective-field-theory regime. Towards the deep IR, this fixed point then moves along a separatrix into the free fixed point at $\alpha_N=0, \alpha_y=0$,  or alternatively is driven towards the strongly-coupled regime if $\alpha_y$ is allowed to deviate from zero. In principle, it is of course also conceivable that the asymptotically safe matter-gravity fixed point lies at finite $\alpha_y, \alpha_N$, and the flow towards the IR then passes by the interacting gauge-Yukawa fixed point and enters the strongly coupled regime at large $\alpha_N$ in the deep IR. 

\subsection{Next-to-next-to leading order}
\begin{figure}[!t]
\begin{center}
\includegraphics[width=0.5\linewidth]{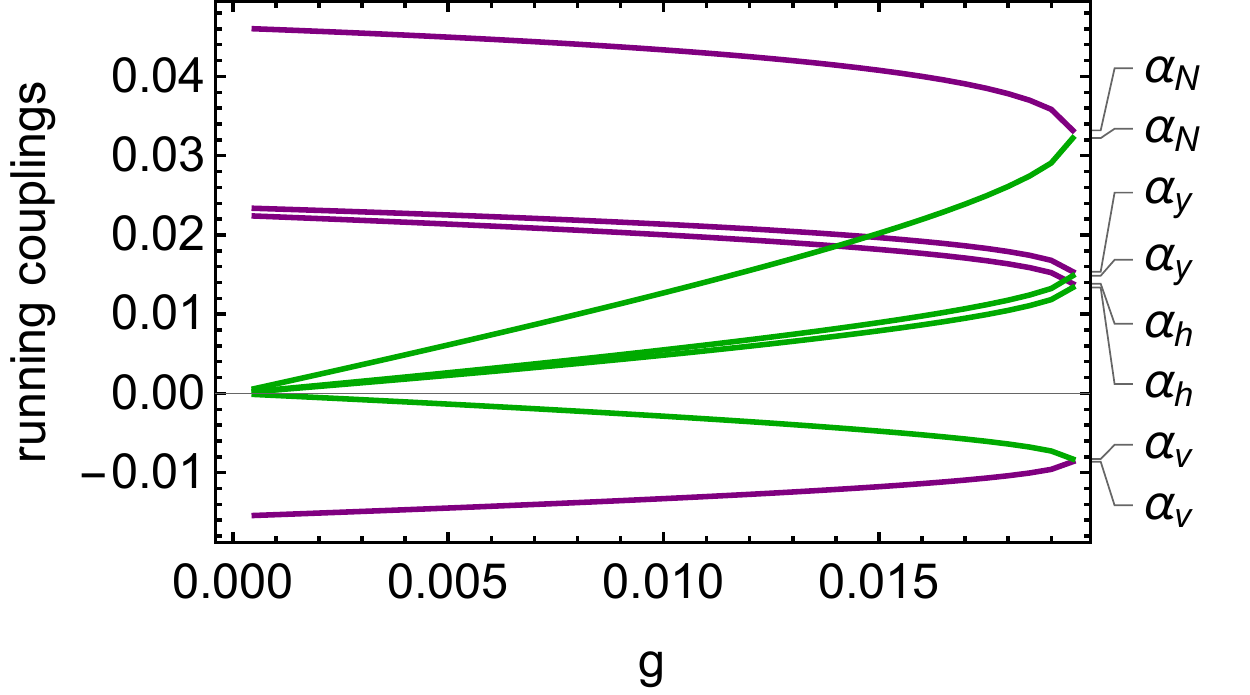}
\end{center}
\caption{\label{fig:FPanniNNLO} Fixed-point values in NNLO at the UV fixed point (purple) and the IR fixed point (green) as a function of $g$ for $\epsilon=0.1$.}
\end{figure}

\begin{figure}[!t]
\includegraphics[width=0.45\linewidth]{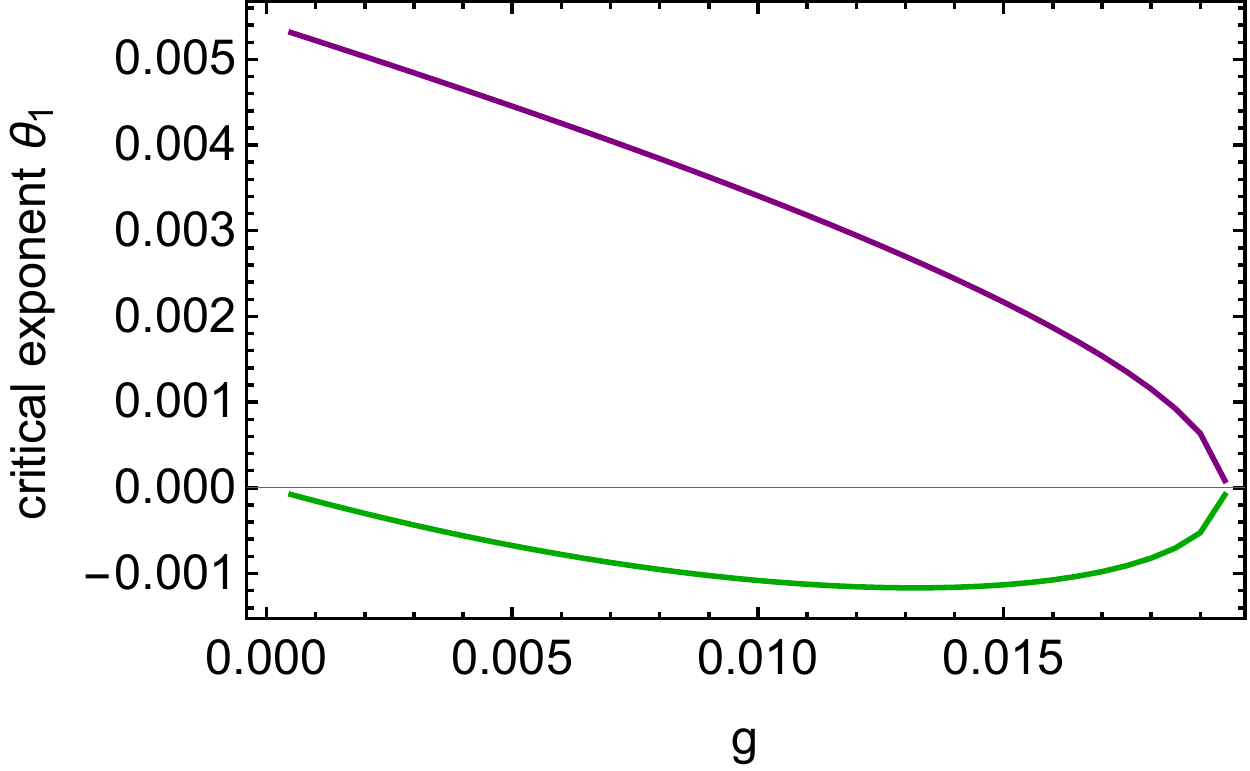}\quad
\includegraphics[width=0.45\linewidth]{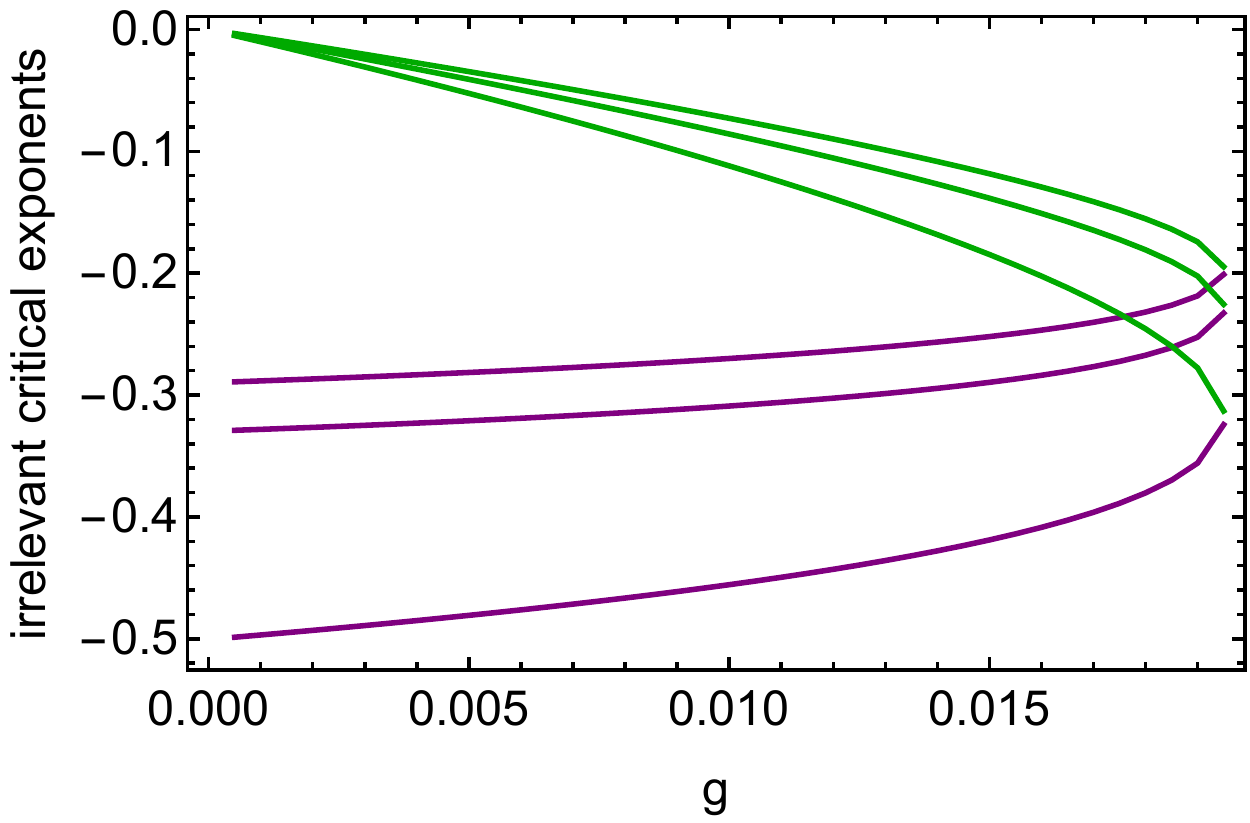}
\caption{\label{fig:thetatab_NNLO} We show the critical exponents in NNLO at the UV fixed point (purple) and the IR fixed point (green) as a function of $g$ for $\epsilon=0.1$. The left panel shows that the fully IR attractive and the UV attractive fixed point approach each other and annihilate at the point where $\theta_1=0$. }
\end{figure}

At NNLO in the beta functions the scalar 
self-interactions come into play and contribute to the gauge-Yukawa sector, as discussed in \cite{Litim:2014uca}.
For a $N_F \times N_F$ complex matrix scalar, two quartic terms exist, namely
\be
L_U =  u {\rm Tr} \left(H^{\dagger}H \right)^2, \quad L_V =  v \left({\rm Tr} H^{\dagger}H \right)^2,
\ee
where we have already made the transition to the Euclidean setting. 
Note that these Lagrange densities are independent of the metric, thus the 
coupling of these operators to metric fluctuations arises through the metric 
determinant and hence the gravity contributions to these 
scalar interactions are identical. 
In the gauge $\alpha =0, \beta=1$, we find the following result
\be
\beta_{u}\Big|_{\rm grav} = u\, \frac{3}{\pi}g \,\,\,\,\,\,\, , \,\,\,\,\,\,\, 
\beta_{v}\Big|_{\rm grav} = v\, \frac{3}{\pi}g.
\ee
Note that the positive sign of the gravity contribution underlies a scenario in which the Higgs mass could become predictable starting from a fully asymptotically safe matter-gravity model \cite{Shaposhnikov:2009pv,Wetterich:2017ixo}.
As the canonical dimensions of $u,v$ are $4-d$, this result can again be read as an effective dimensional increase induced by metric fluctuations. For the couplings
\be
\alpha_h = \frac{u N_F}{(4 \pi)^2}, \quad \alpha_v = \frac{v N_F^2}{(4\pi^2)},
\ee
this implies exactly the same gravity contribution.
Stability of the potential  is given 
if 
\be
\alpha_{h\, \ast} + \alpha_{v\, \ast}>0.
\ee
 For matter theories without gravity, vacuum stability has been discussed in \cite{Litim:2015iea}.
The  UV fixed point from the next-to-leading-order 
analysis splits into two fixed points (at $g=0$), of which only one satisfies 
the stability criterion. Similar to the behavior at 
next-to-leading order, under the effect of gravity the Gau\ss{}ian fixed point 
again splits into several fixed points, one of which satisfies all criteria for 
viability and becomes a fully IR- attractive fixed point, 
cf.~Fig.~\ref{fig:thetatab_NNLO}. 
At a critical value of $g$, it collides with the UV fixed point and they move 
off into the complex plane, cf.~Fig.~\ref{fig:FPanniNNLO}. In 
summary, 
the qualitative effects of graviton fluctuations on the 
gauge-Yukawa system do not appear to change when next-to-next-to-leading 
order terms in the $\beta$-functions of the matter sector are taken into 
account.

\subsection{Impact of matter on the running of $g$}

Until now, we have worked under the assumption that an effective-field-theory regime for quantum gravity exists in which the Planck scale does not run, i.e., the dimensionless Newton coupling runs quadratically with scale. We will now point out that this assumption might require further study in the Veneziano limit: Quantum fluctuations of matter impact the running of the Newton coupling \cite{Larsen:1995ax,Calmet:2008df,Dona:2013qba} according to
\be
\beta_g =2 g - \frac{g^2}{6 \pi}\left(N_S + 2 N_D - 4 N_V \right),\label{eq:running_quad}
\ee
where for the case of interest here the number of scalars $N_S =N_F^2$, the number of Dirac fermions $N_D = N_F\, N_C$ and the number of vectors is $N_V=N_C^2-1$. In Eq.~\eqref{eq:running_quad}, we have neglected the graviton loop contributions. As we take the Veneziano limit, the term quadratic in $g$ starts to feature a divergent coefficient. Matter fluctuations thus drive a fast running of the Newton coupling. Using Eq.~\eqref{eq:def_epsilon}, we write
\be
\beta_g = 2 g+ \frac{g^2}{6\pi} N_C^2\left(\left(\epsilon +\frac{11}{2}\right)^2 +2\left(\epsilon +\frac{11}{2}\right)- 4+\frac{4}{N_C^2}  \right).
\ee
Accordingly, we introduce 
\be
\tilde{g} = g\, N_C^2
\ee
This provides us with a beta function for the rescaled Newton coupling
\be
\beta_{\tilde{g}}= 2 \tilde{g} + \frac{\tilde{g}^2}{6\pi} \left(\frac{149}{4}+13 \epsilon + \epsilon^2 \right).
\ee
We observe that matter fluctuations result in an accelerated running of the Newton coupling. On the other hand, the rescaling of the Newton coupling leads to a decoupling of graviton effects from the running of the matter couplings. In combination, we conclude that this regime requires further studies, and presumably a fully-fledged quantum gravity treatment. 

\section{Summary and conclusions}
We explore the leading-order quantum- gravity effects on fixed points in semi-simple gauge theories and gauge-Yukawa theories within the framework of the functional 
RG. Our analysis pertains to the effective-field-theory regime for gravity, i.e., we make no assumptions about the UV completion for gravity, and explore the leading-order quantum gravity effects only.  We assume that for this study it is sufficient to work with a constant Planck mass, i.e., with a dimensionless Newton coupling that runs canonically.
The range of validity of our analysis does therefore not extend beyond the Planck scale, where higher-order quantum gravity effects must be taken into account.\\
Quantum-gravity 
 contributions to the beta functions of the couplings in the system are linear in those couplings and in the dimensionless Newton coupling $g$, i.e., they act like corrections to the \emph{dimensional} scaling of the system. In particular, we observe that within the functional RG framework the quantum-gravity correction to the running of the gauge couplings can be re-interpreted as a running gauge coupling in a \emph{dimensionally reduced} setting. This has several important consequences: Firstly, it implies that gauge theories which are asymptotically free remain so under the impact of quantum gravity. Therefore, the phase structure of semi-simple gauge theories as explored in \cite{Esbensen:2015cjw}, which feature IR
as well as potential UV fixed points, remains unaffected by quantum gravity. The stabilizing effect of quantum gravity fluctuations even extends the region in parameter space in $(N,M)$, where fully and partially interacting fixed points exist. The values of all fixed points depend on the dimensionless Newton coupling and exhibit a corresponding scale-dependence, unless the dimensionless Newton coupling is held fixed.

Secondly, while the quantum-gravity effect on the Yukawa coupling and the scalar self-interaction takes the form of a lowered scaling dimension, i.e., of ``dimensional increase", the  ``dimensional reduction" of the gauge system through quantum gravity is decisive for the altered dynamics of the gauge-Yukawa system.  It is intriguing that a dimensional reduction, typically exhibited by the spectral dimension has also been observed in the deep quantum regime in many quantum gravity approaches. Of course the scaling dimensions of particular matter couplings are conceptually different from the spectral dimension. Nevertheless one might speculate whether quantum gravity induces a dimensional reduction in some form in order to become UV complete, and the observed dimensional reduction in the gauge-Yukawa system is an imprint of this property that survives in the effective-field theory regime.
 The effective dimensional reduction induces a split of the IR- attractive free fixed point into several fixed points, one of which is interacting and totally  
IR- attractive. As a function of the dimensionless Newton coupling $g$, the interacting 
 IR fixed point moves towards the interacting  UV fixed point which is already present in the gravity-free system. These two fixed points collide and annihilate. In other words, quantum-gravity effects can destroy the fixed-point structure in the system as they trigger a fixed-point collision.  In the presence of $g$, all but the free fixed point in the system depend on $g$. If canonical scaling is assumed for $g$, all fixed points therefore become parametrically scale dependent, realizing a particular form of universal behavior.
For any value of $\epsilon$, which controls the approach to the Veneziano limit for $\epsilon \rightarrow 0$, there is a critical value of the dimensionless Newton coupling $g_{\rm crit}$, such that this fixed-point collision occurs. Even as this occurs, another partially interacting fixed point is induced by graviton fluctuations and is left behind. We conclude 
that quantum gravity effects might have a significant impact on the fixed-point structure in these models,  and can trigger fixed-point destructions.

\emph{Acknowledgements}\\
We thank D.~F.~Litim,  J.~M.~Pawlowski and F.~Sannino for insightful discussions. This work has been supported by the DFG under the Emmy Noether program, grant no.~Ei-1037-1. A.~Held also acknowledges support by the Studienstiftung des deutschen Volkes.

\appendix
\section{Gauge dependence}
\label{sec:gaugedependence}
We treat metric fluctuations 
around a flat background with a split according to
\be
g_{\mu \nu} = \delta_{\mu\nu} +\sqrt{32\pi \frac{k^2}{M_{\rm Planck}^2}}\, 
h_{\mu\nu} \,.
\ee
To obtain the propagator of metric fluctuations, we start from an 
Einstein-Hilbert action at vanishing cosmological constant, supplemented by a 
harmonic-type gauge condition, 
\bea
S&=& S_{\rm EH}+ S_{\rm gf}\\
&=& \frac{-1}{16\pi G} \int d^4x \sqrt{g}R + \frac{1}{2 \alpha}\int d^4x \left( 
\partial^{\mu}h_{\mu\nu} - 
\frac{1+\beta}{4}\partial_{\nu}h\right)\delta^{\nu\lambda}\left( 
\partial^{\kappa}h_{\kappa\lambda} - 
\frac{1+\beta}{4}\partial_{\lambda}h\right).\nonumber
\eea

\begin{figure}[t!]
\includegraphics[width=0.45\linewidth]{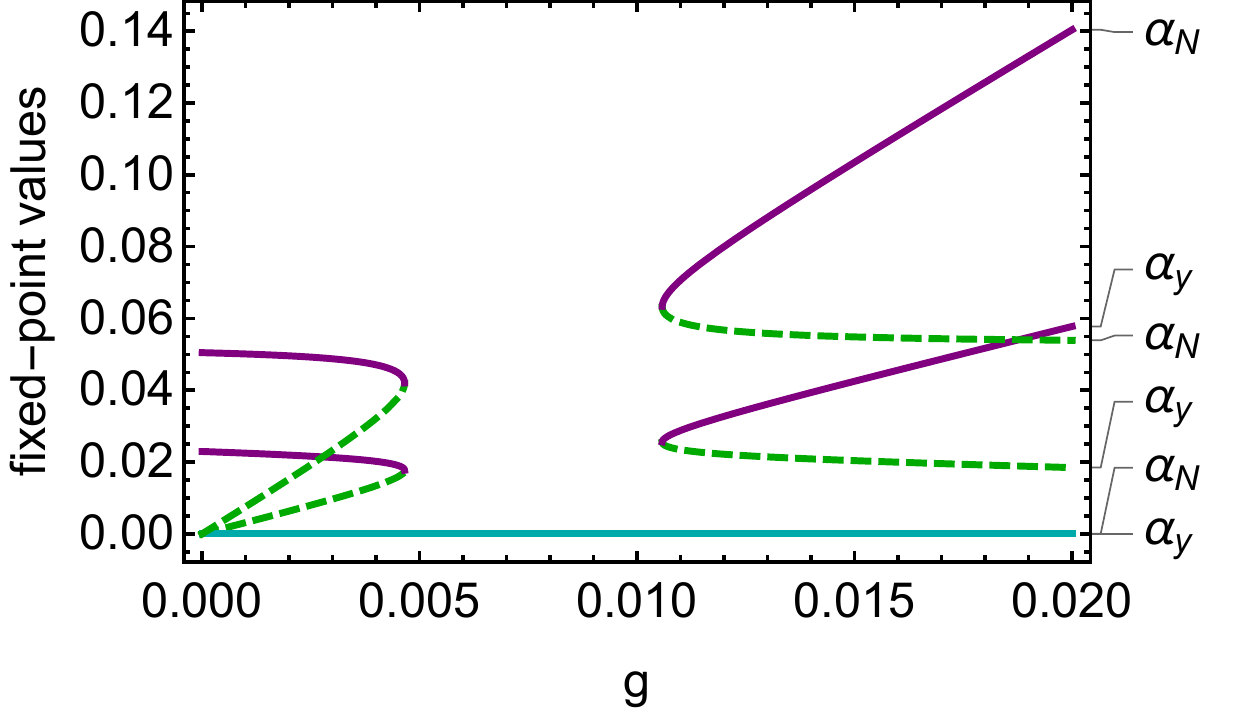}\quad
\includegraphics[width=0.45\linewidth]{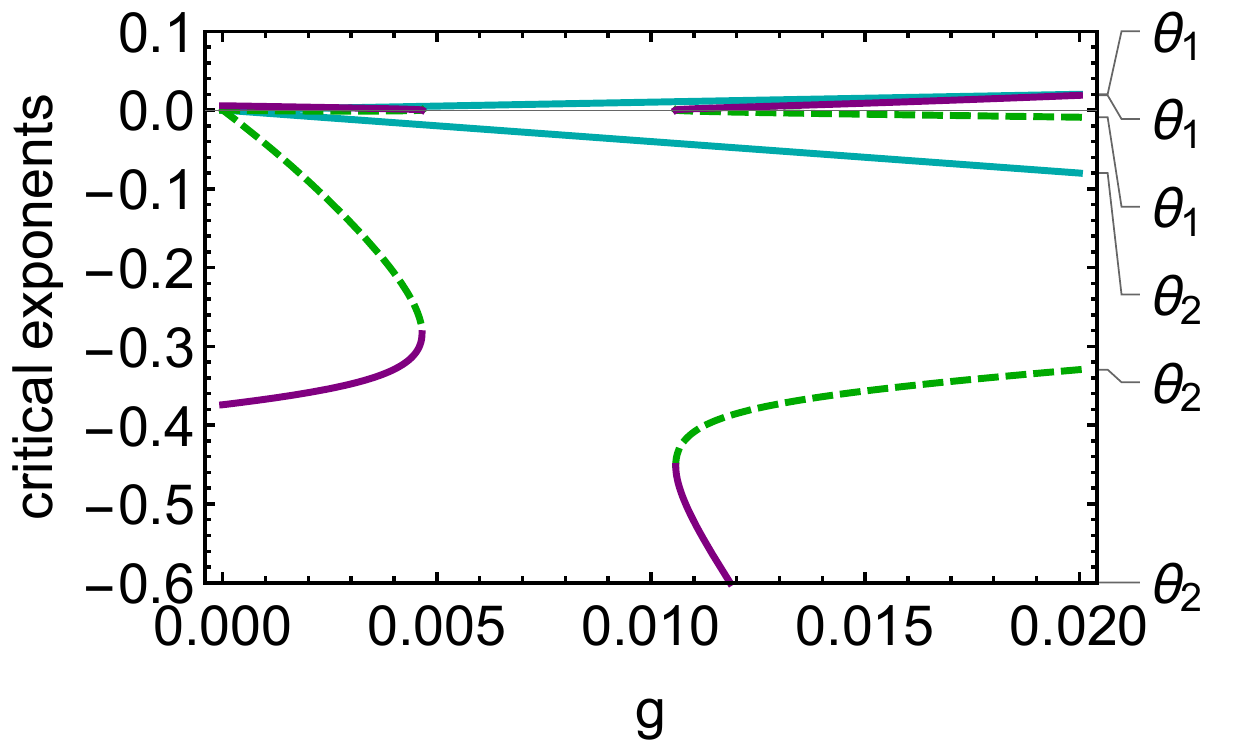}\\
\includegraphics[width=0.45\linewidth]{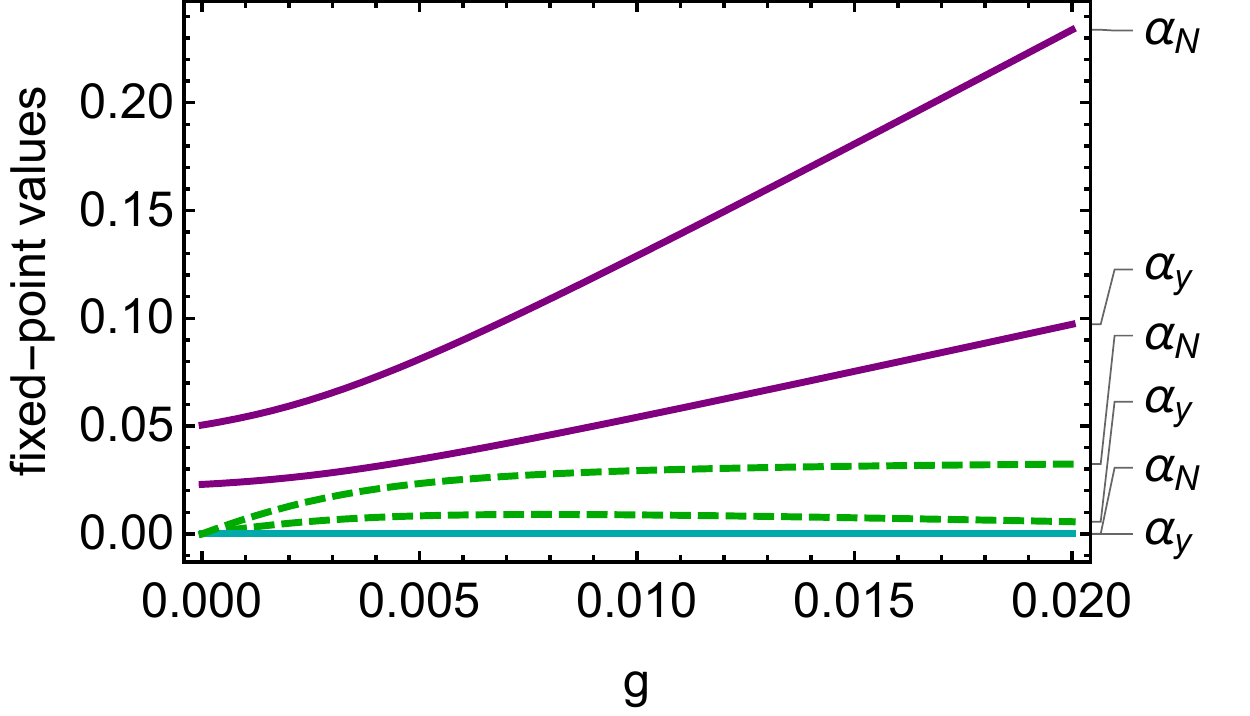}\quad
\includegraphics[width=0.45\linewidth]{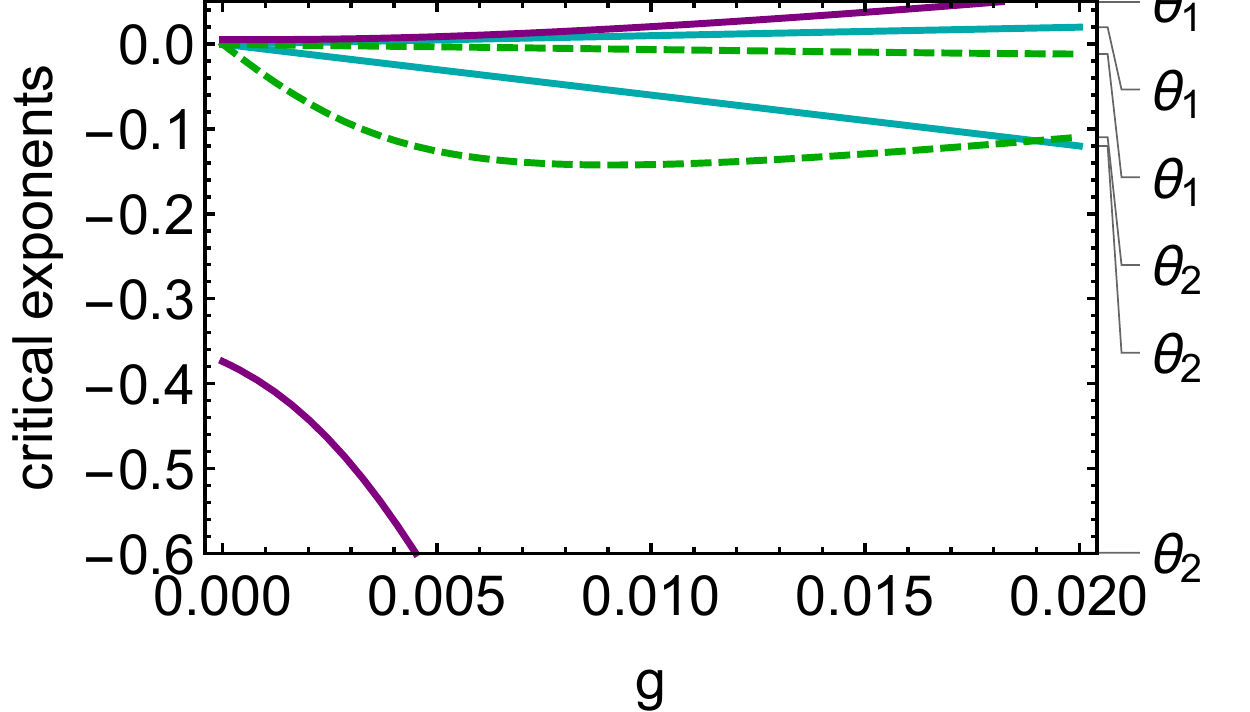}\\
\caption{\label{fig:FPnoanniplots} We show fixed-point values and critical 
exponents for $\mathcal{D}^{\mathrm{grav}}_{\alpha_N}=1$ and 
$\mathcal{D}^{\mathrm{grav}}_{\alpha_y} 
=1$ (upper panels) and $\mathcal{D}^{\mathrm{grav}}_{\alpha_y} =1.5$ (lower panels).}
\end{figure}

\begin{figure}[!t]
\begin{center}
\includegraphics[width=0.45\linewidth,clip=true, trim=10cm 8cm 10cm 5cm]{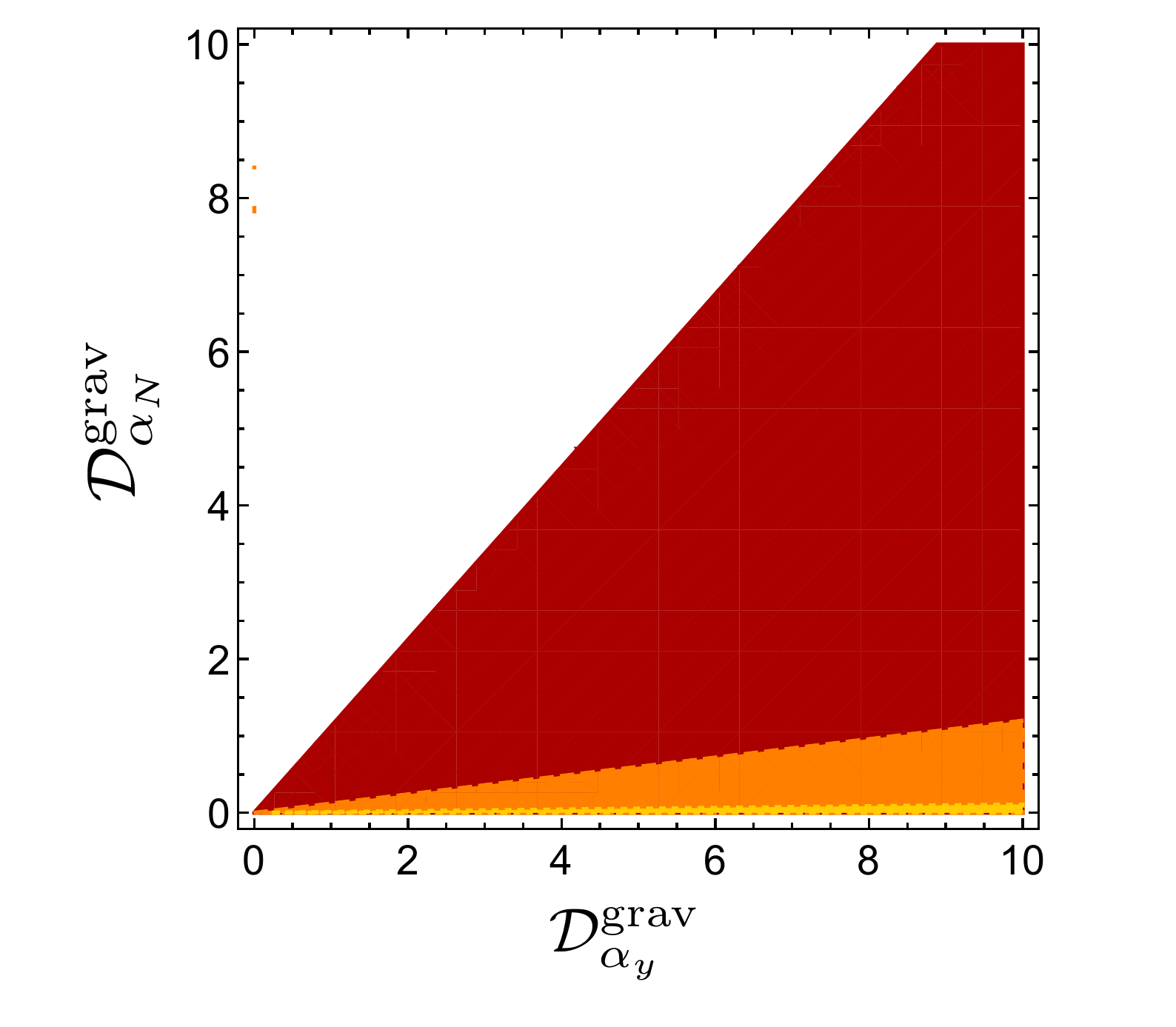}
\end{center}
\caption{\label{fig:Cplot}  We show the region where the fixed-point collision is avoided for $\epsilon=0.5$ in red, $\epsilon=0.1$ in 
orange, $\epsilon=0.01$ in yellow.}
\end{figure}

If we parameterize quantum-gravity corrections to the beta functions of the 
system in the form
\bea
\beta_{\alpha_N}\Big|_{\rm grav}&=&-\mathcal{D}^{\mathrm{grav}}_{\alpha_N}\, g\, 
\alpha_N,\\
\beta_{\alpha_y}\Big|_{\rm grav}&=& \mathcal{D}^{\mathrm{grav}}_{\alpha_y}\, g\, \alpha_y,
\eea
with $\mathcal{D}^{\mathrm{grav}}_{\alpha_y}>0,\, \mathcal{D}^{\mathrm{grav}}_{\alpha_N}>0$,
then we observe that for relatively large $\epsilon$, certain regions in the 
space of $\mathcal{D}^{\mathrm{grav}}_{\alpha_N}$ and $\mathcal{D}^{\mathrm{grav}}_{\alpha_y}$ do 
not lead to a fixed-point annihilation, cf.~Fig.~\ref{fig:FPnoanniplots}, and 
Fig.~\ref{fig:Cplot}. As $g$ increases, an intriguing dynamics 
is triggered
in the matter sector: 
Depending on the relative size of the gravity-contribution to the flow of 
$\alpha_N$ versus the flow of $\alpha_y$, the interacting UV fixed point of the 
matter sector 
could run
towards larger or smaller values as a function of $g$. In 
the meantime, the newly emerged interacting IR fixed point runs towards larger 
values. Thus, these two fixed points either move towards each other while $g$ 
increases, or both move to larger values. In the former case, 
there is a critical value $g_{\rm crit}$ at which fixed-point annihilation 
occurs, while 
there is no special value of $g$ in the latter case.
Thus it is crucial to 
explore whether 
the dynamics of the second case
can be 
reached by a choice of gauge.

We obtain the following gauge-dependent gravity contributions to the beta 
functions, where we emphasize that $\alpha$ is a gauge parameter here, 
not to be confused with the gauge coupling,
\bea
-\mathcal{D}_{\alpha_N}^{\rm grav} &=& -\frac{(15-10 \beta + \beta^2 +\alpha(11-6 \beta + 
\beta^2))}{3 \pi (\beta-3)^2} \, ,
\\
	\mathcal{D}^{\rm grav}_{\alpha_y} &=&- \frac{ \left(\alpha  \left(3 \beta ^2-18 
\beta +31\right)-3 (\beta -1)^2\right)}{3 \pi  (\beta -3)^2}
	-6\frac{ \left(\alpha  \left(5 \beta ^2-30 \beta +53\right)+5 \beta 
^2-50 \beta +81\right)}{16 \pi  (\beta -3)^2}
   \nonumber\\
   &{}&
	-\frac{ (\alpha  (\beta -3)+2 (\beta +3)) 6}{40 \pi  (\beta -3)}
   +\frac{3  (\alpha  (\beta -3)-3 (\beta +13))7}{280 \pi 
 (\beta -3)}
   \nonumber\\
	&{}&+
	\frac{  \left(5 \alpha  \left(5 \beta ^2-30 \beta +53\right)+35 \beta 
^2-258 \beta +339\right)}{10 \pi  (\beta -3)^2} \,.
\eea

\subsection{Next-to leading order}
\begin{figure}[h]
\begin{center}
\includegraphics[width=0.5\linewidth]{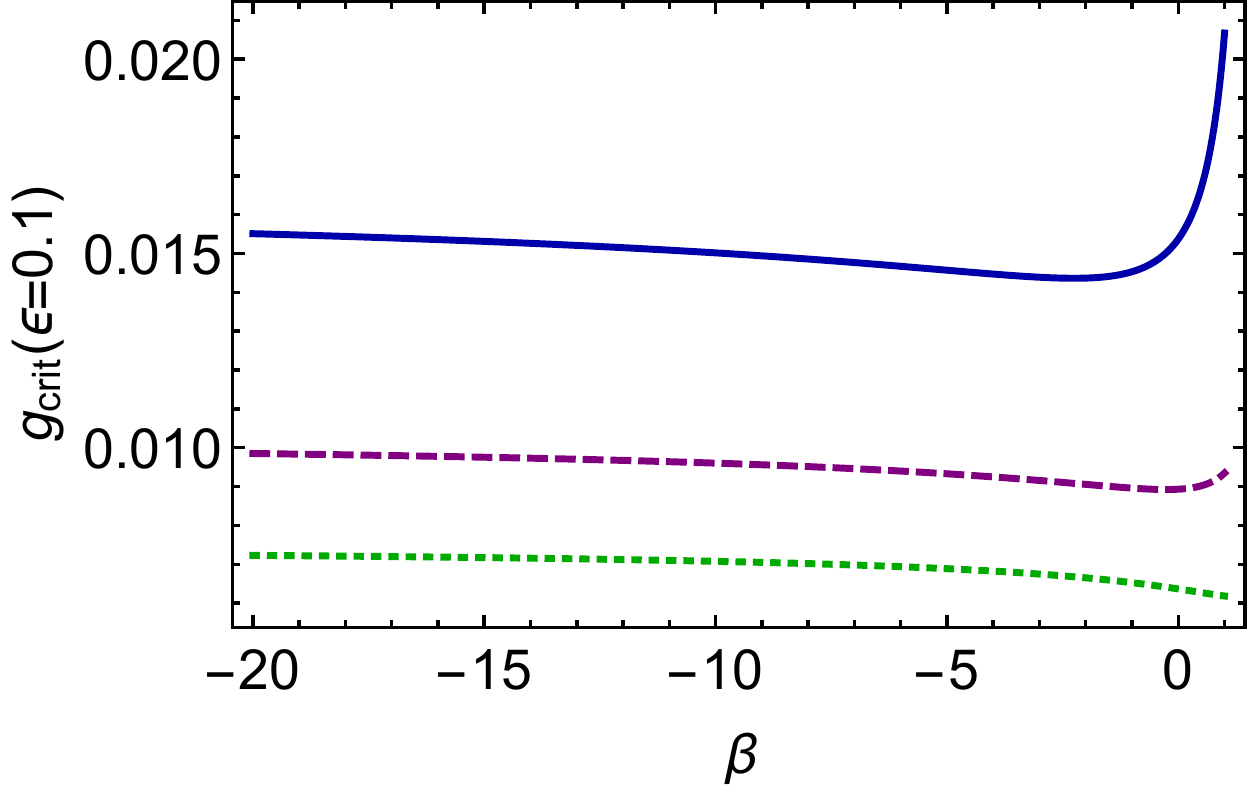}
\end{center}
\caption{\label{fig:gcritbeta}We show the dependence of $g_{\rm crit}$ 
on $\beta$ for $\epsilon=1/10$  for different values of 
$\alpha$: $\alpha=0$ (continuous blue line), $\alpha=0.5$ (purple dashed line) 
and $\alpha=1$ (green dotted line).}
\end{figure}
\begin{figure}[h]
\begin{center}
\includegraphics[width=0.5\linewidth]{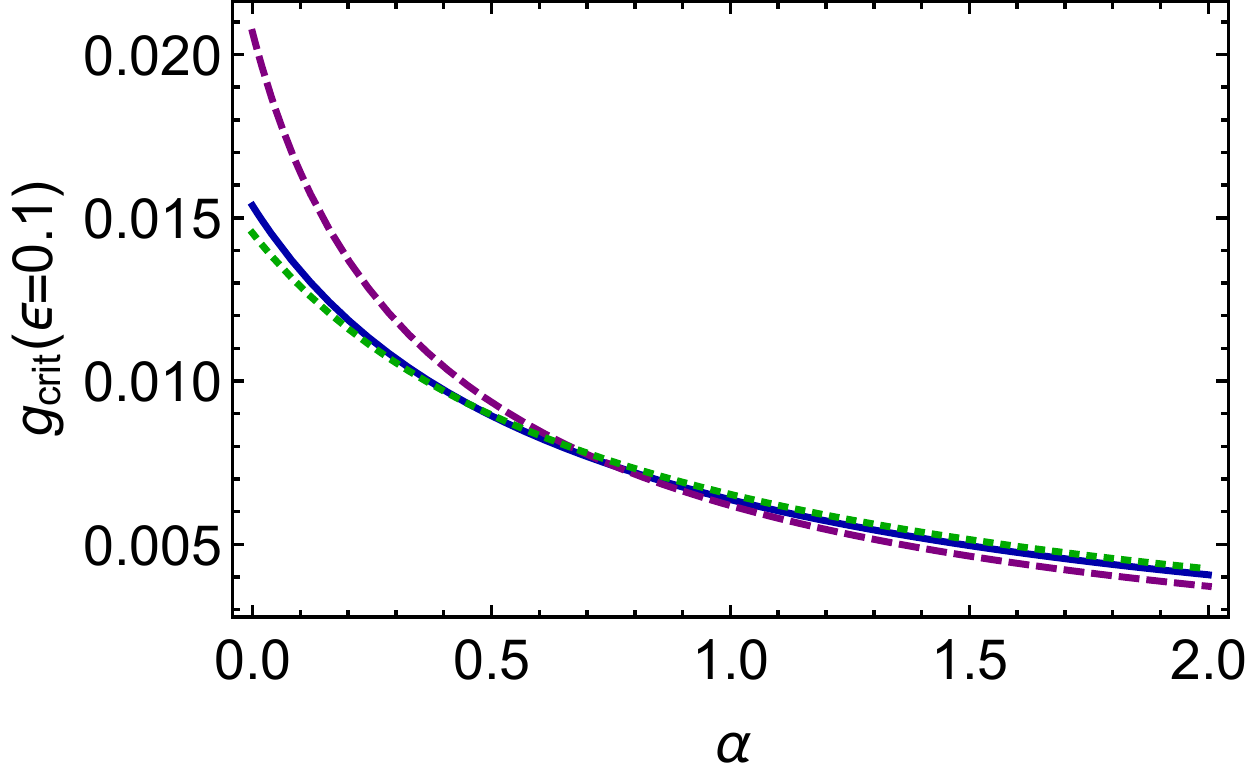}
\end{center}
\caption{\label{fig:gcritalpha}We show the dependence of $g_{\rm crit}$ 
on $\alpha$ for $\epsilon=1/10$ for different values of 
$\beta$: $\beta=0$ 
(continuous blue line), $\beta=1$ (purple dashed line) and $\beta=-1$ (green 
dotted line).}
\end{figure}

When the gauge parameter $\alpha$ is varied, we find 
the opposite behavior and larger values of $\alpha$ 
shift the critical value where the fixed point turns complex to smaller 
values. This behavior is shown in figure \ref{fig:gcritalpha}. 
We are now in position to analyze the gauge dependence of the fixed points. 
From the equations above one can infer the well-known pole of the gravitational 
contributions at $\beta=3$.
As a consequence,
the results cannot be trusted in the vicinity 
of $\beta =3$ due 
to the instability and we restrict ourselves
to $\beta  \lessapprox 1$. 
The general result for the gauge dependence of $g_{\rm 
crit}$ is given by
\bea
g_{\rm crit}&=& \frac{160\pi}{(11+2\epsilon)^4 
\left(-6(140+\beta(\beta-73))+\alpha(263+27\beta(\beta-6)) \right)^2} \cdot 
\nonumber\\
&{}& \cdot \Bigl[ -2(\beta-3)^2(13+2 \epsilon) \Bigl( -1140 
(15+(\beta-10)\beta) + (64620 + \beta(1283 \beta-35699))\epsilon \nonumber\\
&{}& \quad \quad\quad \quad \quad \quad\quad\quad\quad \quad+4 (5220 + 
\beta(73\beta-2809))\epsilon^2 
+12(140+\beta(\beta-73))\epsilon^3\Bigr)\nonumber\\
&{}&\,\,\,+ \alpha (\beta-3)^2 
(13+2\epsilon)\Bigl(2280(11+\beta(\beta-6))+(11583+1427 
\beta(\beta-6))\epsilon\nonumber\\
&{}& \quad \quad\quad \quad \quad \quad\quad\quad\quad \quad + 
4(2013+217\beta(\beta-6))\epsilon^2 +4(263+ 27 \beta(\beta-6))\epsilon^3 
\Bigr)\nonumber\\
&{}& - 4 \sqrt{5} \Bigl\{\Bigl(-(\beta-3)^4(15+\beta(\beta-10) + \alpha(11+ 
\beta(\beta-6))) (13+2 \epsilon)^2 \cdot \nonumber\\
&{}& \quad \quad \quad \quad \cdot(-57+46 \epsilon+ 8 \epsilon^2) \Bigl(1140 
(15+\beta(\beta-10) + \alpha(11+\beta(\beta-6)))\nonumber\\
&{}& \quad \quad \quad \quad \quad \quad + (\alpha(21703+2347\beta(\beta-6)) -2 
(57720 + \beta(823 \beta-31099))) \epsilon \nonumber\\
&{}&\quad \quad \quad \quad \quad \quad + 4 (-9840 + 2\beta(2609-53 \beta) + 
\alpha(2453+257\beta(\beta-6))) \epsilon^2 \nonumber\\
&{}&\quad \quad \quad \quad \quad \quad + 4(-6(140+\beta(\beta-73)) + 
\alpha(263+27\beta(\beta-6)))\epsilon^3
\Bigr)
 \Bigr)
\Bigr\}^{\frac{1}{2}}
 \Bigr]
\eea

The main message of this analysis can be summarized as follows.
The qualitative features of the fixed point structure are independent of $\beta 
\in (-\infty,1]$ and $0 < \alpha < \infty$.   
For fixed $\alpha$, the fixed point annihilation is shifted towards smaller 
values of $g$ as $\beta$ is lowered and converges (for $\epsilon= 1/10$) 
towards

\begin{equation}
\lim_{\beta \rightarrow -\infty} g_{\mathrm{crit}} = \frac{55 \pi}{11586+40 
\sqrt{83603}} \approx 0.007 \,,  
\end{equation}
with corresponding fixed point values
\begin{equation}
\lim_{\beta \rightarrow -\infty} \alpha_{N,*} =  
\frac{55(-545+2\sqrt{83603}}{74774} \approx 0.02 \,,
\end{equation}
and
\begin{equation}
\lim_{\beta \rightarrow -\infty} \alpha_{y,*} =  
\frac{5(-2573163+9580\sqrt{83603}}{87934224} \approx 0.01 \,.
\end{equation}
For fixed $\alpha$ the parametric dependence on $\beta$ is 
shown in figure \ref{fig:gcritbeta}. For any epsilon, there is value at which 
$g_{\mathrm{crit}}$ disappears, i.e.\ where no fixed point annihilation takes 
place. However, this is typically in a region with large $\beta$. Moreover, one 
can see that there is a minimum at some $\alpha$-dependent value of 
$g_{\mathrm{crit}}$ as a function 
of $\beta$, but there is no discontinuity or any further peculiar 
behavior of the curve all the way down to the smooth limit $\beta 
\rightarrow -\infty$.


\begin{thebibliography}{99}

\bibitem{Esbensen:2015cjw} 
  J.~K.~Esbensen, T.~A.~Ryttov and F.~Sannino,
  Phys.\ Rev.\ D {\bf 93}, no. 4, 045009 (2016)
  doi:10.1103/PhysRevD.93.045009
  [arXiv:1512.04402 [hep-th]].
  
\bibitem{Litim:2014uca} 
  D.~F.~Litim and F.~Sannino,
  JHEP {\bf 1412}, 178 (2014)
  doi:10.1007/JHEP12(2014)178
  [arXiv:1406.2337 [hep-th]].

\bibitem{Weinberg:1980gg}
  S.~Weinberg,
{\it  In *Hawking, S.W., Israel, W.: General Relativity*, 790-831}
(Cambridge University Press, Cambridge, 1980).


\bibitem{Gross:1973id} 
  D.~J.~Gross and F.~Wilczek,
  Phys.\ Rev.\ Lett.\  {\bf 30}, 1343 (1973).
  doi:10.1103/PhysRevLett.30.1343


\bibitem{Politzer:1973fx} 
  H.~D.~Politzer,
  Phys.\ Rev.\ Lett.\  {\bf 30}, 1346 (1973).
  doi:10.1103/PhysRevLett.30.1346



\bibitem{Peskin1980}
Peskin, Michael E.,
Phys. Lett. B94 (1980)
doi:10.1016/0370-2693(80)90848-5

  
\bibitem{Gies:2003ic} 
  H.~Gies,
  Phys.\ Rev.\ D {\bf 68}, 085015 (2003)
  doi:10.1103/PhysRevD.68.085015
  [hep-th/0305208].

%
\bibitem{Fei:2014yja} 
  L.~Fei, S.~Giombi and I.~R.~Klebanov,
  Phys.\ Rev.\ D {\bf 90}, no. 2, 025018 (2014)
  doi:10.1103/PhysRevD.90.025018
  [arXiv:1404.1094 [hep-th]].
  
\bibitem{Gracey:2015tta} 
  J.~A.~Gracey,
  Phys.\ Rev.\ D {\bf 92}, no. 2, 025012 (2015)
  doi:10.1103/PhysRevD.92.025012
  [arXiv:1506.03357 [hep-th]].
  
\bibitem{Eichhorn:2016hdi} 
  A.~Eichhorn, L.~Janssen and M.~M.~Scherer,
  Phys.\ Rev.\ D {\bf 93}, no. 12, 125021 (2016)
  doi:10.1103/PhysRevD.93.125021
  [arXiv:1604.03561 [hep-th]].
  
  
\bibitem{Gawedzki:1985ed} 
  K.~Gawedzki and A.~Kupiainen,
  Phys.\ Rev.\ Lett.\  {\bf 55}, 363 (1985).
  doi:10.1103/PhysRevLett.55.363
  
\bibitem{Hands:1992be} 
  S.~Hands, A.~Kocic and J.~B.~Kogut,
  Annals Phys.\  {\bf 224}, 29 (1993)
  doi:10.1006/aphy.1993.1039
  [hep-lat/9208022].
  
\bibitem{Braun:2010tt} 
  J.~Braun, H.~Gies and D.~D.~Scherer,
  Phys.\ Rev.\ D {\bf 83}, 085012 (2011)
  doi:10.1103/PhysRevD.83.085012
  [arXiv:1011.1456 [hep-th]].
  
    %
\bibitem{Bond:2016dvk} 
  A.~D.~Bond and D.~F.~Litim,
  arXiv:1608.00519 [hep-th].
  
\bibitem{Bond:2017sem} 
  A.~Bond and D.~Litim,
  PoS LATTICE {\bf 2016}, 208 (2017).
  
  
\bibitem{Intriligator:2015xxa} 
  K.~Intriligator and F.~Sannino,
  JHEP {\bf 1511}, 023 (2015)
  doi:10.1007/JHEP11(2015)023
  [arXiv:1508.07411 [hep-th]].
 
  
\bibitem{Pelaggi:2017wzr} 
  G.~M.~Pelaggi, F.~Sannino, A.~Strumia and E.~Vigiani,
  arXiv:1701.01453 [hep-ph].
  
\bibitem{Bond:2017wut} 
  A.~D.~Bond, G.~Hiller, K.~Kowalska and D.~F.~Litim,
  arXiv:1702.01727 [hep-ph].
  
    %
\bibitem{Sannino:2014lxa} 
  F.~Sannino and I.~M.~Shoemaker,
  Phys.\ Rev.\ D {\bf 92}, no. 4, 043518 (2015)
  doi:10.1103/PhysRevD.92.043518
  [arXiv:1412.8034 [hep-ph]].
  
\bibitem{Nielsen:2015una} 
  N.~G.~Nielsen, F.~Sannino and O.~Svendsen,
  Phys.\ Rev.\ D {\bf 91}, 103521 (2015)
  doi:10.1103/PhysRevD.91.103521
  [arXiv:1503.00702 [hep-ph]].
  
\bibitem{Rischke:2015mea} 
  D.~H.~Rischke and F.~Sannino,
  Phys.\ Rev.\ D {\bf 92}, no. 6, 065014 (2015)
  doi:10.1103/PhysRevD.92.065014
  [arXiv:1505.07828 [hep-th]].
  
\bibitem{Bajc:2016efj} 
  B.~Bajc and F.~Sannino,
  JHEP {\bf 1612}, 141 (2016)
  doi:10.1007/JHEP12(2016)141
  [arXiv:1610.09681 [hep-th]].

  

  
 
\bibitem{ASgravity}
  M.~Reuter,
  Phys.\ Rev.\ D {\bf 57}, 971 (1998)
  [hep-th/9605030];
  M.~Reuter and F.~Saueressig,
  Phys.\ Rev.\ D {\bf 65}, 065016 (2002)
  [hep-th/0110054];
  D.~F.~Litim,
  Phys.\ Rev.\ Lett.\  {\bf 92}, 201301 (2004)
  [hep-th/0312114];
  A.~Codello, R.~Percacci and C.~Rahmede,
  Annals Phys.\  {\bf 324}, 414 (2009)
  [arXiv:0805.2909 [hep-th]];
  D.~Benedetti, P.~F.~Machado and F.~Saueressig,
  Mod.\ Phys.\ Lett.\ A {\bf 24}, 2233 (2009)
  [arXiv:0901.2984 [hep-th]];
    D.~Benedetti and F.~Caravelli,
  JHEP {\bf 1206}, 017 (2012)
  [Erratum-ibid.\  {\bf 1210}, 157 (2012)]
  [arXiv:1204.3541 [hep-th]];
  N.~Christiansen, D.~F.~Litim, J.~M.~Pawlowski and A.~Rodigast,
  Phys.\ Lett.\ B {\bf 728}, 114 (2014)
  [arXiv:1209.4038 [hep-th]];
  J.~A.~Dietz and T.~R.~Morris,
  JHEP {\bf 1301}, 108 (2013)
  doi:10.1007/JHEP01(2013)108
  [arXiv:1211.0955 [hep-th]];
    K.~Falls, D.~F.~Litim, K.~Nikolakopoulos and C.~Rahmede,
  arXiv:1301.4191 [hep-th];
    D.~Becker and M.~Reuter,
  Annals Phys.\  {\bf 350}, 225 (2014)
  doi:10.1016/j.aop.2014.07.023
  [arXiv:1404.4537 [hep-th]];
      N.~Christiansen, B.~Knorr, J.~Meibohm, J.~M.~Pawlowski and M.~Reichert,
  Phys.\ Rev.\ D {\bf 92}, no. 12, 121501 (2015)
  doi:10.1103/PhysRevD.92.121501
  [arXiv:1506.07016 [hep-th]];
    H.~Gies, B.~Knorr, S.~Lippoldt and F.~Saueressig,
  Phys.\ Rev.\ Lett.\  {\bf 116}, no. 21, 211302 (2016)
  doi:10.1103/PhysRevLett.116.211302
  [arXiv:1601.01800 [hep-th]];
  N.~Christiansen,
  arXiv:1612.06223 [hep-th];
    T.~Denz, J.~M.~Pawlowski and M.~Reichert,
  arXiv:1612.07315 [hep-th];
  K.~Falls,
  arXiv:1702.03577 [hep-th].
  
  \bibitem{ASgravitymatter}
  G.~Narain and R.~Percacci,
  Class.\ Quant.\ Grav.\  {\bf 27}, 075001 (2010)
  doi:10.1088/0264-9381/27/7/075001
  [arXiv:0911.0386 [hep-th]];
  A.~Eichhorn,
  Phys.\ Rev.\ D {\bf 86}, 105021 (2012)
  doi:10.1103/PhysRevD.86.105021
  [arXiv:1204.0965 [gr-qc]];
  T.~Henz, J.~M.~Pawlowski, A.~Rodigast and C.~Wetterich,
  Phys.\ Lett.\ B {\bf 727}, 298 (2013)
  doi:10.1016/j.physletb.2013.10.015
  [arXiv:1304.7743 [hep-th]];
    P.~Don\`a, A.~Eichhorn and R.~Percacci,
  Phys.\ Rev.\ D {\bf 89}, no. 8, 084035 (2014)
  doi:10.1103/PhysRevD.89.084035
  [arXiv:1311.2898 [hep-th]];
  R.~Percacci and G.~P.~Vacca,
  Eur.\ Phys.\ J.\ C {\bf 75}, no. 5, 188 (2015)
  doi:10.1140/epjc/s10052-015-3410-0
  [arXiv:1501.00888 [hep-th]];
    J.~Meibohm, J.~M.~Pawlowski and M.~Reichert,
  Phys.\ Rev.\ D {\bf 93}, no. 8, 084035 (2016)
  doi:10.1103/PhysRevD.93.084035
  [arXiv:1510.07018 [hep-th]];
    P.~Don\`a, A.~Eichhorn, P.~Labus and R.~Percacci,
  Phys.\ Rev.\ D {\bf 93}, no. 4, 044049 (2016)
  Erratum: [Phys.\ Rev.\ D {\bf 93}, no. 12, 129904 (2016)]
  doi:10.1103/PhysRevD.93.129904, 10.1103/PhysRevD.93.044049
  [arXiv:1512.01589 [gr-qc]];
    J.~Biemans, A.~Platania and F.~Saueressig,
  arXiv:1702.06539 [hep-th].
  
     \bibitem{ASreviews}
  M.~Niedermaier and M.~Reuter,
  Living Rev.\ Rel.\  {\bf 9}, 5 (2006);
  M.~Niedermaier,
  Class.\ Quant.\ Grav.\  {\bf 24}, R171 (2007)
  [gr-qc/0610018];
  R.~Percacci,
  In Oriti, D. (ed.): ``Approaches to quantum gravity'' 111-128
  [arXiv:0709.3851 [hep-th]];
  D.~F.~Litim,
  arXiv:0810.3675 [hep-th];
  D.~F.~Litim,
  Phil.\ Trans.\ Roy.\ Soc.\ Lond.\ A {\bf 369}, 2759 (2011)
  [arXiv:1102.4624 [hep-th]];
  R.~Percacci,
  arXiv:1110.6389 [hep-th];
  M.~Reuter and F.~Saueressig,
  New J.\ Phys.\  {\bf 14}, 055022 (2012)
  [arXiv:1202.2274 [hep-th]];
  M.~Reuter and F.~Saueressig,
  arXiv:1205.5431 [hep-th];
%
  S.~Nagy,
 Annals Phys.\  {\bf 350}, 310 (2014)
  doi:10.1016/j.aop.2014.07.027
  [arXiv:1211.4151 [hep-th]];
    A.~Ashtekar, M.~Reuter and C.~Rovelli,
  arXiv:1408.4336 [gr-qc];
  A.~Bonanno and F.~Saueressig,
  arXiv:1702.04137 [hep-th].

  \bibitem{EFTabelian}
  S.~P.~Robinson and F.~Wilczek,
  Phys.\ Rev.\ Lett.\  {\bf 96}, 231601 (2006)
  doi:10.1103/PhysRevLett.96.231601
  [hep-th/0509050];
  A.~R.~Pietrykowski,
  Phys.\ Rev.\ Lett.\  {\bf 98}, 061801 (2007)
  doi:10.1103/PhysRevLett.98.061801
  [hep-th/0606208];
  D.~Ebert, J.~Plefka and A.~Rodigast,
  Phys.\ Lett.\ B {\bf 660}, 579 (2008)
  doi:10.1016/j.physletb.2008.01.037
  [arXiv:0710.1002 [hep-th]];
  D.~J.~Toms,
  Phys.\ Rev.\ D {\bf 76}, 045015 (2007)
  doi:10.1103/PhysRevD.76.045015
  [arXiv:0708.2990 [hep-th]];
  D.~J.~Toms,
  Phys.\ Rev.\ Lett.\  {\bf 101}, 131301 (2008)
  doi:10.1103/PhysRevLett.101.131301
  [arXiv:0809.3897 [hep-th]];
  D.~J.~Toms,
  Phys.\ Rev.\ D {\bf 80}, 064040 (2009)
  doi:10.1103/PhysRevD.80.064040
  [arXiv:0908.3100 [hep-th]];
  D.~J.~Toms,
  Nature {\bf 468}, 56 (2010)
  doi:10.1038/nature09506
  [arXiv:1010.0793 [hep-th]];
  D.~J.~Toms,
  Phys.\ Rev.\ D {\bf 84}, 084016 (2011).
  doi:10.1103/PhysRevD.84.084016
  
\bibitem{Daum:2009dn} 
  J.~E.~Daum, U.~Harst and M.~Reuter,
  JHEP {\bf 1001}, 084 (2010)
  doi:10.1007/JHEP01(2010)084
  [arXiv:0910.4938 [hep-th]].
  
  \bibitem{Folkerts:2011jz} 
  S.~Folkerts, D.~F.~Litim and J.~M.~Pawlowski,
  Phys.\ Lett.\ B {\bf 709}, 234 (2012)
  doi:10.1016/j.physletb.2012.02.002
  [arXiv:1101.5552 [hep-th]].
  
\bibitem{Harst:2011zx} 
  U.~Harst and M.~Reuter,
  JHEP {\bf 1105}, 119 (2011)
  doi:10.1007/JHEP05(2011)119
  [arXiv:1101.6007 [hep-th]].
  
\bibitem{Christiansen:2017gtg} 
  N.~Christiansen and A.~Eichhorn,
  arXiv:1702.07724 [hep-th].
  
\bibitem{Wetterich:1992yh} 
  C.~Wetterich,
  Phys.\ Lett.\ B {\bf 301}, 90 (1993).
  
        %
\bibitem{Morris:1993qb} 
  T.~R.~Morris,
  Int.\ J.\ Mod.\ Phys.\ A {\bf 9}, 2411 (1994)
  [hep-ph/9308265].
  
  \bibitem{Papenbrock}
  T.~Papenbrock and C.~Wetterich,
  Z.\ Phys.\ C {\bf 65}, 519 (1995).

\bibitem{Litim:2001ky} 
  D.~F.~Litim and J.~M.~Pawlowski,
  Phys.\ Rev.\ D {\bf 65}, 081701 (2002)
  doi:10.1103/PhysRevD.65.081701
  [hep-th/0111191].

\bibitem{Reuter:1993kw} 
  M.~Reuter and C.~Wetterich,
  Nucl.\ Phys.\ B {\bf 417}, 181 (1994).
  doi:10.1016/0550-3213(94)90543-6
  
\bibitem{Gies:2002af} 
  H.~Gies,
  Phys.\ Rev.\ D {\bf 66}, 025006 (2002)
  doi:10.1103/PhysRevD.66.025006
  [hep-th/0202207].
  
    %
\bibitem{Canet:2003qd} 
  L.~Canet, B.~Delamotte, D.~Mouhanna and J.~Vidal,
  Phys.\ Rev.\ B {\bf 68}, 064421 (2003)
  doi:10.1103/PhysRevB.68.064421
  [hep-th/0302227].
  
\bibitem{Litim:2010tt} 
  D.~F.~Litim and D.~Zappala,
  Phys.\ Rev.\ D {\bf 83}, 085009 (2011)
  doi:10.1103/PhysRevD.83.085009
  [arXiv:1009.1948 [hep-th]].
  
    \bibitem{Berges:2000ew}
  J.~Berges, N.~Tetradis and C.~Wetterich,
  Phys.\ Rept.\  {\bf 363} (2002) 223
  [hep-ph/0005122].
  
  \bibitem{Polonyi:2001se}
  J.~Polonyi,
  Central Eur.\ J.\ Phys.\  {\bf 1}, 1 (2003) 
 [hep-th/0110026].
%
\bibitem{Pawlowski:2005xe}
  J.~M.~Pawlowski,
  Annals Phys.\  {\bf 322} (2007) 2831 
  [arXiv:hep-th/0512261].
%

\bibitem{Gies:2006wv}
  H.~Gies, Lect.\ Notes Phys.\ {\bf 852}, 287 (2012)
  [arXiv:hep-ph/0611146]. 

\bibitem{Delamotte:2007pf} 
  B.~Delamotte,
  Lect.\ Notes Phys.\  {\bf 852}, 49 (2012)
  [cond-mat/0702365].
  
%
\bibitem{Rosten:2010vm}
  O.~J.~Rosten,
  arXiv:1003.1366 [hep-th].
  
\bibitem{Braun:2011pp} 
  J.~Braun,
  J.\ Phys.\ G {\bf 39}, 033001 (2012)
  [arXiv:1108.4449 [hep-ph]].

   
\bibitem{Litim:2001up} 
  D.~F.~Litim,
  Phys.\ Rev.\ D {\bf 64}, 105007 (2001)
  doi:10.1103/PhysRevD.64.105007
  [hep-th/0103195].

\bibitem{Carlip:2016qrb} 
  S.~Carlip,
  Int.\ J.\ Mod.\ Phys.\ D {\bf 25}, no. 12, 1643003 (2016)
  doi:10.1142/S0218271816430033
  [arXiv:1605.05694 [gr-qc]].

%
\bibitem{Ambjorn:2005db} 
  J.~Ambjorn, J.~Jurkiewicz and R.~Loll,
  Phys.\ Rev.\ Lett.\  {\bf 95}, 171301 (2005)
  doi:10.1103/PhysRevLett.95.171301
  [hep-th/0505113].

\bibitem{Lauscher:2005qz} 
  O.~Lauscher and M.~Reuter,
  JHEP {\bf 0510}, 050 (2005)
  doi:10.1088/1126-6708/2005/10/050
  [hep-th/0508202].
  
\bibitem{Horava:2009if} 
  P.~Horava,
  Phys.\ Rev.\ Lett.\  {\bf 102}, 161301 (2009)
  doi:10.1103/PhysRevLett.102.161301
  [arXiv:0902.3657 [hep-th]].

\bibitem{Calcagni:2013vsa} 
  G.~Calcagni, A.~Eichhorn and F.~Saueressig,
  Phys.\ Rev.\ D {\bf 87}, no. 12, 124028 (2013)
  doi:10.1103/PhysRevD.87.124028
  [arXiv:1304.7247 [hep-th]].

\bibitem{Carlip:2015mra} 
  S.~Carlip,
  Class.\ Quant.\ Grav.\  {\bf 32}, no. 23, 232001 (2015)
  doi:10.1088/0264-9381/32/23/232001
  [arXiv:1506.08775 [gr-qc]].
  

  

  
  
\bibitem{Veneziano:1979ec} 
  G.~Veneziano,
  Nucl.\ Phys.\ B {\bf 159}, 213 (1979).
  doi:10.1016/0550-3213(79)90332-8

\bibitem{Zanusso:2009bs} 
  O.~Zanusso, L.~Zambelli, G.~P.~Vacca and R.~Percacci,
  Phys.\ Lett.\ B {\bf 689}, 90 (2010)
  doi:10.1016/j.physletb.2010.04.043
  [arXiv:0904.0938 [hep-th]].
  
  
\bibitem{Vacca:2010mj} 
  G.~P.~Vacca and O.~Zanusso,
  Phys.\ Rev.\ Lett.\  {\bf 105}, 231601 (2010)
  doi:10.1103/PhysRevLett.105.231601
  [arXiv:1009.1735 [hep-th]].
  
  %
\bibitem{Eichhorn:2016esv} 
  A.~Eichhorn, A.~Held and J.~M.~Pawlowski,
  Phys.\ Rev.\ D {\bf 94}, no. 10, 104027 (2016)
  doi:10.1103/PhysRevD.94.104027
  [arXiv:1604.02041 [hep-th]].
  
\bibitem{Oda:2015sma} 
  K.~y.~Oda and M.~Yamada,
  Class.\ Quant.\ Grav.\  {\bf 33}, no. 12, 125011 (2016)
  doi:10.1088/0264-9381/33/12/125011
  [arXiv:1510.03734 [hep-th]].
  
\bibitem{Hamada:2017rvn} 
  Y.~Hamada and M.~Yamada,
  arXiv:1703.09033 [hep-th].
  
  \bibitem{Held:2017}
  A.~Eichhorn and A.~Held,
  to appear.
  
\bibitem{Codello:2016muj} 
  A.~Codello, K.~Langæble, D.~F.~Litim and F.~Sannino,
  JHEP {\bf 1607}, 118 (2016)
  doi:10.1007/JHEP07(2016)118
  [arXiv:1603.03462 [hep-th]].
  
  
  
\bibitem{Dona:2013qba} 
  P.~Don\`a , A.~Eichhorn and R.~Percacci,
  Phys.\ Rev.\ D {\bf 89}, no. 8, 084035 (2014)
  doi:10.1103/PhysRevD.89.084035
  [arXiv:1311.2898 [hep-th]].
  
\bibitem{Meibohm:2015twa} 
  J.~Meibohm, J.~M.~Pawlowski and M.~Reichert,
  Phys.\ Rev.\ D {\bf 93}, no. 8, 084035 (2016)
  doi:10.1103/PhysRevD.93.084035
  [arXiv:1510.07018 [hep-th]].
  
  
  
\bibitem{Litim:2015iea} 
  D.~F.~Litim, M.~Mojaza and F.~Sannino,
  JHEP {\bf 1601}, 081 (2016)
  doi:10.1007/JHEP01(2016)081
  [arXiv:1501.03061 [hep-th]].
  
  
\bibitem{Shaposhnikov:2009pv} 
  M.~Shaposhnikov and C.~Wetterich,
  Phys.\ Lett.\ B {\bf 683}, 196 (2010)
  doi:10.1016/j.physletb.2009.12.022
  [arXiv:0912.0208 [hep-th]].

\bibitem{Wetterich:2017ixo} 
  C.~Wetterich,
  arXiv:1704.08040 [gr-qc].

\bibitem{Larsen:1995ax} 
  F.~Larsen and F.~Wilczek,
  Nucl.\ Phys.\ B {\bf 458}, 249 (1996)
  doi:10.1016/0550-3213(95)00548-X
  [hep-th/9506066].

\bibitem{Calmet:2008df} 
  X.~Calmet, S.~D.~H.~Hsu and D.~Reeb,
  Phys.\ Rev.\ Lett.\  {\bf 101}, 171802 (2008)
  doi:10.1103/PhysRevLett.101.171802
  [arXiv:0805.0145 [hep-ph]].







  
\end{thebibliography}
\end{document}